\pdfoutput=1
\documentclass[journal]{vgtc}                %

\ifpdf%
  \pdfoutput=1\relax                   %
  \usepackage{graphicx}                %
  \DeclareGraphicsExtensions{.pdf,.png,.jpg,.jpeg} %
\else%
  \usepackage{graphicx}                %
  \DeclareGraphicsExtensions{.eps}     %
\fi%

\graphicspath{{figures/}{pictures/}{images/}{./}} %

\usepackage{microtype}                 %
\PassOptionsToPackage{warn}{textcomp}  %
\usepackage{textcomp}                  %
\usepackage{mathptmx}                  %
\usepackage{times}                     %
\usepackage{cite}                      %
\usepackage{tabu}                      %
\usepackage{booktabs}                  %

\usepackage{tabularx}

\usepackage{array}
\newcolumntype{L}[1]{>{\raggedright\let\newline\\\arraybackslash\hspace{0pt}}m{#1}}
\newcolumntype{C}[1]{>{\centering\let\newline\\\arraybackslash\hspace{0pt}}m{#1}}
\newcolumntype{R}[1]{>{\raggedleft\let\newline\\\arraybackslash\hspace{0pt}}m{#1}} 

\usepackage{enumitem}
\setlist{nolistsep}

\usepackage{subfigure}

\usepackage{textcomp}
\usepackage{color,soul}
\usepackage[dvipsnames]{xcolor}
\usepackage{todonotes}

\usepackage{soul}

\usepackage[final]{changes}

\definecolor{fifth}{HTML}{C7E9C0}
\definecolor{fourth}{HTML}{A1D99B}
\definecolor{third}{HTML}{74C476}
\definecolor{second}{HTML}{31A354}
\definecolor{first}{HTML}{006D2C}

\definecolor{second_fourth}{HTML}{BAE4B3}
\definecolor{third_first}{HTML}{238B45}
\definecolor{first_conf}{HTML}{08519C}
\definecolor{last_conf}{HTML}{C6DBEF}

\onlineid{0}

\vgtccategory{Research}
\vgtcpapertype{evaluation}

\title{Origin-Destination Flow Maps in Immersive Environments}

\author{Yalong Yang, Tim Dwyer, Bernhard Jenny, Kim Marriott, Maxime Cordeil and Haohui Chen}
\authorfooter{
\item
Yalong Yang is with Monash University and Data61, CSIRO, Australia. E-mail: yalong.yang@monash.edu.
\item
Tim Dwyer, Bernhard Jenny, Kim Marriott and Maxime Cordeil are with Monash University. E-mail: \{tim.dwyer, bernie.jenny, kim.marriott, max.cordeil\}@monash.edu.
\item
Haohui Chen is with Data61, CSIRO, Australia. E-mail: caronhaohui.chen@data61.csiro.au.
}

\abstract{
Immersive virtual- and augmented-reality headsets can overlay a flat image against any surface or hang virtual objects in the space around the user. The technology is rapidly improving and may, in the long term, replace traditional flat panel displays in many situations. When displays are no longer intrinsically flat, how should we use the space around the user for abstract data visualisation? In this paper, we ask this question with respect to origin-destination flow data in a global geographic context. We report on the findings of three studies exploring different spatial encodings for flow maps. The first experiment focuses on different 2D and 3D encodings for flows on flat maps. We find that participants are significantly more accurate with raised flow paths whose height is proportional to flow distance but fastest with traditional straight line 2D flows. In our second and third experiment we compared flat maps, 3D globes and a novel interactive design we call \emph{MapsLink}, involving a pair of linked flat maps. We find that participants took significantly more time with MapsLink than other flow maps while the 3D globe with raised flows was  the fastest, most accurate, and most preferred method. 
Our work suggests that \emph{careful} use of the third spatial dimension can resolve visual clutter in complex flow maps.
} %


\CCScatlist{ %
 \CCScat{K.6.1}{Management of Computing and Information Systems}%
{Project and People Management}{Life Cycle};
 \CCScat{K.7.m}{The Computing Profession}{Miscellaneous}{Ethics}
}

\teaser{
\setlength{\belowcaptionskip}{0cm}
  \centering
  \includegraphics[height=4cm]{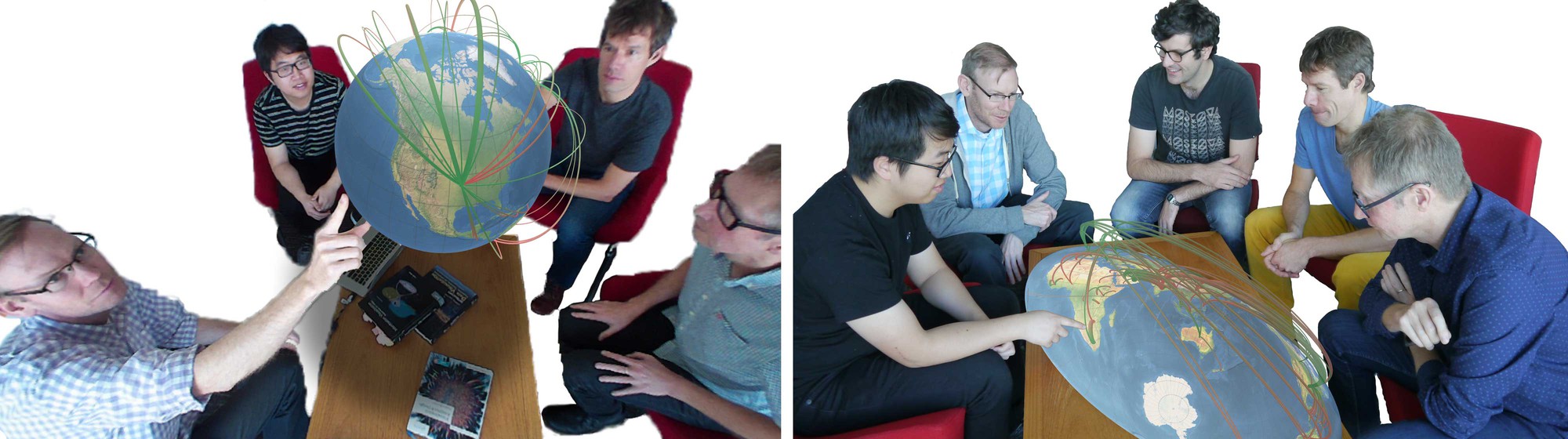}
  \caption{3D Globe (left) was the fastest and most accurate of our tested visualisations.
  Flat maps with curves of height proportional to distance (right) were more accurate but slower than 2D straight lines.
  Experiments were conducted \added{individually} in virtual reality but our motivation for this work is to support future mixed-reality \added{collaborative} scenarios
  like those envisaged in these figures.}
  \label{fig:teaser}
}

\vgtcinsertpkg

\begin{document}

\hyphenation{Maps-Link}
\firstsection{Introduction}
\maketitle

In many applications it is important to visualise the flows between different geographic locations.  Such flows include, for example, migration patterns \cite{Tobler:1987kn}, 
movement of goods or knowledge~\cite{paci2009knowledge}, disease or animals~\cite{gilbert2005cattle,guo2007visual}. Here we restrict attention to flows that start and end at fixed geographic locations but whose exact trajectory is either unknown or irrelevant.
Visualisation of such origin-destination flows is a difficult information visualisation challenge, because both the locations of origins and destinations and the connections between them need to be represented. 
The most common visualisation is the OD (origin-destination) flow map, where each flow is represented as a line connecting the origin and destination on a map. A disadvantage of OD flow maps is that they become cluttered and difficult to read as the number of flows increases. Nonetheless, flow maps remain popular as they are intuitive and well-suited to showing a small number of flows.

With the arrival of commodity head-mounted displays (HMDs) for virtual-reality (VR), e.g. HTC Vive, and  augmented-reality (AR), e.g. Microsoft Hololens, Meta2 and Magic Leap, we can expect to see geographic visualisations such as OD flow maps used in mixed-reality (MR) applications. Such applications include situated analytics~\cite{elsayed2016situated} where visualisations can be made available in almost any environment such as in the field, surgery or factory floor, and collaborative visualisation scenarios, where two or more people wearing HMDs can each see and interact with visualisations while still seeing each other~\cite{Cordeil:2016io}.

The key question we address is whether traditional 2D OD flow maps are the best way to show origin-destination flow in such immersive environments or whether some variant that makes use of a third dimension may be better.  
While current guidelines for information visualisation design caution against the use of 3D spatial encodings of abstract data~\cite[Ch.\ 6]{munzner2014visualization}, in the case of global flow data visualised in immersive environments the third dimension offers an extended design space that is appealing for a number of reasons:
\begin{itemize}[leftmargin=1em]
\item The height dimension offers the possibility of an additional spatial encoding for data attributes. %
\item Lifting flow curves off the map may reduce clutter and provide better visibility of the underlying map.
\item In immersive environments, occlusions can be resolved by natural head movements or gesture manipulations to change the view angle.
\item In 2D flow maps the flows may be perceived as trajectories (highways, shipping routes, etc.), lifting them into the third dimension may resolve this ambiguity.
\end{itemize}
We investigate this design space through three controlled user studies. To the best of our knowledge we are the first to do so. Our paper has three main contributions.

The first contribution is to chart the design space for 3D flow maps (Sec.~\ref{sec:design-space}). Following D\"ubel \emph{et al.}~\cite{dubel20142d} we separate the design space into two orthogonal components: the representation of flow, e.g.\ straight or curved lines in 2 or 3D, and the representation of the geographic reference space, e.g.\ 3D globe or flat map. Furthermore, origins and destinations can either be shown on the same or separate globes or maps. This leads to a rich multi-dimensional design space.

The second contribution is evaluation of different flow maps in VR differing in the representation of flow (Study 1, Sec.~\ref{sec:study1}). We compared  2D flow representations with (a) straight and (b) curved flow lines, and 3D flow tubes with (c) constant height, height varying with (d) quantity and (e) distance between start and end points. We measured time and accuracy to find and compare the magnitude of flow between two pairs of locations. We found that participants were most accurate using 3D flows on flat maps when flow height was proportional to flow distance and that this was the preferred representation. Participants were less accurate with straight 2D flows than with 3D flows, but faster.

The third contribution is evaluation of different flow map visualisations primarily varying in the representation of the reference space (Studies 2 and 3). For the same task as above, we first compared a flat map with 2D flow lines, a flat map with 3D flow tubes, a 3D globe with 3D tubes, and a novel design called \emph{MapsLink} involving a pair of flat maps linked with 3D tubes. We found (Sec.~\ref{sec:study2}) that participants were much slower with the linked map pair than all other representations. What surprised us was that participants were more accurate with the 3D globe than with 2D flows on a flat map and linked map pairs. Participants were also faster with the 3D globe than with 3D flows on a flat map. The final experiment (Sec.~\ref{sec:study3}) was similar but with higher densities of flows. This found the 3D globe to be the fastest and most accurate flow visualisation. It was also the preferred representation.

The performance of the 3D globe is unexpected. While we assumed that 3D flows might reduce the problem of clutter, we did not expect that the 3D globe with its potential shortcomings of occlusion and distortion would be more effective than 2D or 3D flows on a flat map. However, this result accords with \cite{Yang:2018mg}, who found that in VR environments 3D globes were better than maps for a variety of map reading tasks.

\begin{figure*}
	\setlength{\belowcaptionskip}{0cm}
	\centering
	\includegraphics[width=.98\textwidth]{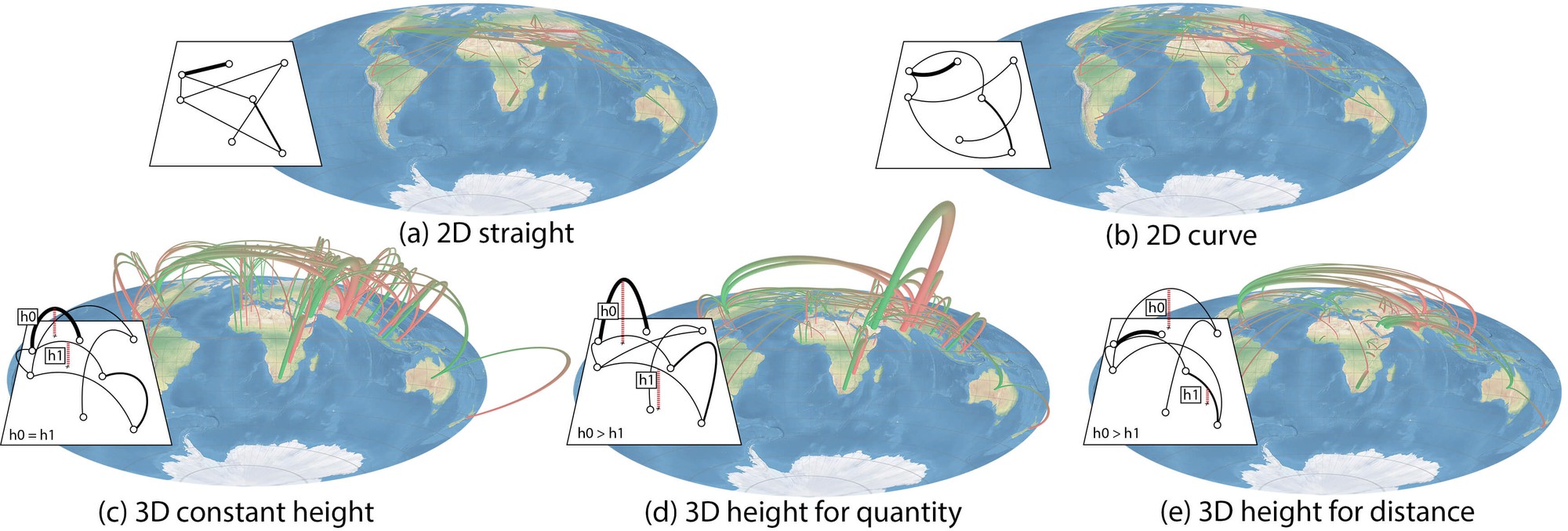}
	\caption{Study 1: Tested 2D and 3D flow maps.}
	\label{fig:first-study-vis}
	\vspace{-2em}
\end{figure*}

\section{Related Work}
\label{sec:relwork}
\subsection{Immersive Analytics}
With the commodification of VR and AR HMDs there is growing interest in how to visualise abstract data in immersive environments~\cite{Immersive15,itoh2016ImmersiveAnalytics,dwyer2016ImmersiveAnalytics,bach2017immersive}. While there is considerable caution about the use of 3D in abstract data visualisation, e.g.\cite[Chap.\ 6]{munzner2014visualization}, there is also a realisation that data visualisation in immersive environments will be increasingly common because of the growing flexibility that MR HMDs offer over traditional desktop environments including ability to associate data with objects in physical environments~\cite{elsayed2016situated} and to support collaboration~\cite{Cordeil:2016io}.
Particular use cases  include 3D graph layout e.g.~\cite{ware2005reevaluating,Kwon:2016go,Huang2017AGesture,Cordeil:2016io} and multivariate data visualisation, e.g.~\cite{cordeil2017imaxes,Butscher:2018ct}.

There seems considerable potential to use the third dimension for visualisation of spatially embedded data~\cite{dubel20142d}. This is because the geographic reference space typically takes up two-dimensions, adding a third dimension  offers the possibility of an additional spatial encoding for data attributes. Indeed common geographic representations such as space-time cubes or prism maps routinely use a third dimension even though they are to be displayed on standard desktop displays. \added{Evaluations of such 3D geographic visualisation (e.g.~\cite{kaya20143d}) on flat displays has not yielded positive results.  However, there has been little research into the effectiveness of such visualisation using modern head-tracked binocular HMDs.} The most relevant study by Yang \textit{et al.}~\cite{Yang:2018mg} compared task performance for three standard map reading tasks\textemdash distance comparison, area comparison and direction estimation\textemdash using different 2D and 3D representations for the Earth. They compared a 3D exocentric globe placed in front of the viewer,  an egocentric 3D globe placed around the viewer, a flat map (rendered to a plane in VR) and a curved map, created by projecting the map onto a section of a sphere curved around the user. In almost all cases the egocentric globe was found to be the least effective visualization and the curved map was generally better than the flat map. Overall the exocentric globe was the best choice.  It was slightly preferred by participants, was more accurate and faster than the other visual presentations for direction estimation and more accurate than the egocentric globe and the flat map for distance comparison, though more time was required for comparison of areas than with flat and curved maps. %

\subsection{2D Flow Visualisation}
Shortly after Henry Drury Harness created the first known flow map in 1837~\cite{Robinson:1955hz}, they were popularised by Minard's work~\cite{Robinson:1967cj}. 
Early digital cartographers extended and improved techniques to map quantitative origin-destination flows and networks using straight lines with width proportionately varying with quantitative attributes  \cite{kern1969Mapit,kadmon1971komplot,wittick1976,Tobler:1981kw,Tobler:1987kn,chicagoAreaTransportStudy}.

Visual clutter is a major concern for maps with a few dozens, hundreds or even thousands of flows. 
Tobler~\cite{Tobler:1987kn} identifies filtering methods and guidelines for simplifying flow data to increase the readability of maps, such as sub-setting (to only show flows of a selected area), thresholding (to only display the largest flows), or merging (to aggregate flows with spatially close origins and destinations).  Interaction, as for example described by van den Elzen and van Wijk~\cite{van2014multivariate}, can provide interactive filtering and aggregation that restricts the set of origins and destinations to a manageable amount.

Research in cartography \cite{koylu2017design, Jenny:2017ci,stephen2017automated} and network visualisation \cite{Holten:2009directed,Holten:2011fp} has identified design principles and aesthetic criteria that can reduce clutter in flow maps.  Jenny \textit{et al.}~\cite{Jenny:2017ci} compiled the following design principles to declutter 2D maps: curving flows, minimising overlap among flows \cite{purchase1995validating} and between flows and nodes \cite{wong2003edgelens}, avoiding acute-angle crossings \cite{huang2014larger}, radially distributing flows \cite{huang2007using}, and stacking small flows on top of large flows \cite{dent2008cartography}. 
Among techniques that employ curvature to improve the readability of flows, bundling of flows into tree-like structures rooted at flow origins has been in use for a long time and several algorithms for their automatic generation have been proposed \cite{Buchin:2011fk,Sun:2018db,Phan:2005cn, debiasi2014supervised,Nocaj:2013gd}. While these techniques offer aesthetically appealing results for one-to-many flow maps, it is unclear how to best adapt them to many-to-many flows.  For example, so-called \emph{confluent} bundling techniques can merge flows without ambiguity but can be challenging to interpret \cite{bach2017towards}.  Another problem with such bundling is that it obscures individual flow lines where they share common end points.  An alternative use of curvature is to fan out (maximise the space between) flows incident to origins or destinations \cite{riche2012exploring}.  Such a technique was recently found to offer improved readability in a controlled study \cite{Jenny:2017ci} and is one of the conditions included in our Study 1 (Sec.~\ref{sec:study1}).

Alternatives to flow maps include density fields to show the spatial concentrations of flows \cite{rae2009spatial,scheepens2011composite}, or OD (origin-destination) maps \cite{wood2010visualisation} that spatially order flows in columns and rows of connectivity matrices embedded on a geographic map. \textit{Flowstrates} \cite{boyandin2011flowstrates} and \textit{MapTrix}~\cite{Yang:2017cy} are two related approaches combining non-geographic OD matrix with two geographic maps for showing the locations of origins and destinations. In \textit{Flowstrates} flows connect a temporal heatmap with the two geographic maps, and in \textit{MapTrix} each flow line links one OD matrix cell with its geographic location on the two connected maps.

\subsection{3D Flow Maps}
Lifting origin-destination flows into the third dimension is not a new idea. Early examples of flow maps drawn with three-dimensional arcs elevated above 2D reference maps and 3D globes were introduced more than 20 years ago \cite{Cox:1995fp, Cox:1996gv, munzner1996visualizing}. The explicit goal of these early maps was to increase readability by disentangling flows.

The height of flows above the reference map or globe can vary with the total volume of flows \cite{Eick96aspectsof}, the distance between endpoints \cite{Cox:1995fp}, the inverse of the distance \cite{Vrotsou:2017im}, time \cite{hagerstrand1970people} or any other attribute \cite{Vrotsou:2017im}. 
Discussion of such 3D ``geovirtual'' environments typically focuses on interactive filtering to deal with clutter (e.g.\ \cite{buschmann2012challenges}).
We are not aware of any study evaluating the effectiveness of these different 3D encodings.

\section{Design Space}
\label{sec:design-space}

Visualisations of OD flow data can present geographic locations of origins and destinations, the direction of flow and flow weight (magnitude or other quantitative attribute). OD flow maps achieve this by showing each flow as a line or arrow on a map connecting the origin and destination. In this section we explore the design space of 3D OD flow maps.  D\"ubel \emph{et al.}~\cite{dubel20142d} categorize geospatial visualizations based on whether the reference space (i.\,e.\ the map or surface) is shown in 2D or 3D and whether the abstract attribute is shown in 2D or 3D. In the case of flow maps,  this categorization implies the design space has two orthogonal components: the representation of flow and the representation of geographic region. 

\subsection{Representation of Flow}
\label{sec:flowrep}
Flow on 2D OD flow maps is commonly shown by a straight line from origin to destination with line width encoding magnitude of flow and an arrowhead showing direction. However, as discussed above, with this encoding visual clutter and line crossings are inevitable, even in small datasets. One way to overcome this is to use curved instead of straight lines, such that the paths are carefully chosen to ``fan out'' or maximise the separation between flows at their origins and destinations \cite{riche2012exploring}. Such curved flow maps have shown to be more effective in ``degree counting'' tasks~\cite{Jenny:2017ci}where the curvature reduced overlaps. Conversely, edge bundling has also been suggested as a way of overcoming clutter. In this approach curved paths are chosen so as to visually combine flows from adjacent regions~\cite{Yang:2017cy}. While bundling can greatly reduce clutter, its disadvantage is that it can make individual flows difficult to follow. Thus, it is probably best suited to overview tasks.

In the case of 3D flow representations, height can be used to
encode flow magnitude or some other quantitative property or to reduce the visual clutter caused by crossings or overlapping. Not only does the use of a third dimension allow flows to be spaced apart, in modern immersive MR environments it also allows the viewer to use motion perspective to better distinguish between flows by either moving their head or by rotating the presentation.
Based on our literature review (Sec.~\ref{sec:relwork}), possible  3D representations include:
\begin{itemize}[leftmargin=1em]
	\item \emph{Constant maximum height:} in which each flow is shown as a curved line connecting the origin and destination and all flows have the same maximum height above the surface of the reference space. This is arguably the simplest way to use height to help address the problem of overlapping flows.
	\item \emph{Height encodes quantity:} height is either proportional or inversely proportional to flow magnitude~\cite{Eick96aspectsof}. This allows double-coding the flow magnitude with both thickness and height.
	\item \emph{Height is proportional to distance:} height is proportional to the distance between origin and destination, short flows will be close to the reference space surface while longer flows will be lifted above it~\cite{Cox:1995fp}. This will tend to vertically separate crossings. 	
	\item \emph{Height is inversely proportional to distance:} this was suggested in~\cite{Vrotsou:2017im}. The advantage is that it increases the visual salience of flows between geographically close locations but at the expense of increasing overlap.  
\end{itemize}

\subsection{Representation of Reference Space}
\label{sec:refspace}

In 2D flow maps the reference space is always a flat map, which can of course also be used in an immersive environment. 
However, in the case of global flows it also makes sense to use a 3D exocentric globe representation in which the flow is shown on a sphere positioned in front of the viewer. The disadvantage of a globe representation is that the curved surface of the globe causes foreshortening and only half of the globe can be seen at one time.  
An alternative is to use a 3D egocentric representation for the globe in which the user is placed inside a large sphere~\cite{Zhang:2016,Yang:2018mg,Zhang:2018jo}. This suffers from similar drawbacks to the exocentric globe: foreshortening and inability to see more than half of the Earth's surface. Furthermore, because of the position of the user it is difficult to see the height of flows. In \cite{Yang:2018mg} we found that the egocentric globe led to worse performance than the exocentric globe for standard map reading tasks and also led to motion sickness. 

Typically the same reference space (map or globe) is used to show both origins and destinations. However, in 2D representations such as \textit{MapTrix} or \textit{Flowstrates} \cite{boyandin2011flowstrates,Yang:2017cy} origins and destinations are shown on different  maps. Potential benefits are reduction of clutter in the reference space representations and clearer depiction of  flow direction. In a 3D environment probably the simplest representation using two reference space representations is to show two 2D maps on flat  planes with  flow shown by connecting  tubes. Such a representation is akin to Collins and Carpendale's \emph{VisLink} technique~\cite{Collins:2007ir}, where multiple 2D abstract data representations are viewed in a 3D environment, with lines linking related points across views.  This was the inspiration for our \emph{MapLink} technique
evaluated in Study 2, Sec.~\ref{sec:study2}.

\section{Study 1: 2D and 3D Flows on Flat Maps}
\label{sec:study1}
The first user study focuses on  representation of flow in VR. It compares readability of  five flow representations (two 2D and three 3D) using  the same reference space representation: a flat map.

\subsection{Visualisations and interactions}
The two 2D representations were:

\noindent\textbf{\textit{2D straight:}} Connecting origins and destinations with straight lines is the most common way to create a 2D flow map (Fig. \ref{fig:first-study-vis}(a)).

\noindent\textbf{\textit{2D curve:}} Using curved flow lines that increase separation and acute angle crossings has been shown to increase readability for dense flows in 2D flow maps~\cite{Jenny:2017ci}. The routing technique in \cite{jenny2017force} was used to created 2D curved flows (Fig. \ref{fig:first-study-vis}(b)).

The three 3D flow representations used 3D tubes to connect origins and destinations. We used a cubic B\'ezier curve to create the tubes using Equation~(\ref{equ:bezier}), where $P_0$ and $P_3$ are the origin and destination of a flow, $0 \le t \le 1$ is the interpolation factor, and $P_1$ and $P_2$ are the two control points to decide the shape of the tube. 
\begin{equation}
	B(t) = (1-t)^3P_0 + 3(1-t)^2tP_1 + 3(1-t)t^2P_2 + t^3P_3
	\label{equ:bezier}
\end{equation}

\noindent\begin{minipage}{0.61\columnwidth}
For $P_1$ and $P_2$, we make their projected positions on 2D map plane the same as $P_0$ (the origin) and $P_3$ (the destination) respectively, so the projected trajectory on the 2D map plane of the 3D tube is a straight line. This allows users to easily follow the direction
\end{minipage}
\begin{minipage}{3.3cm}
\centering
\vspace{0mm}
\small
\fontfamily{phv}\selectfont
\includegraphics[width=3.4cm]{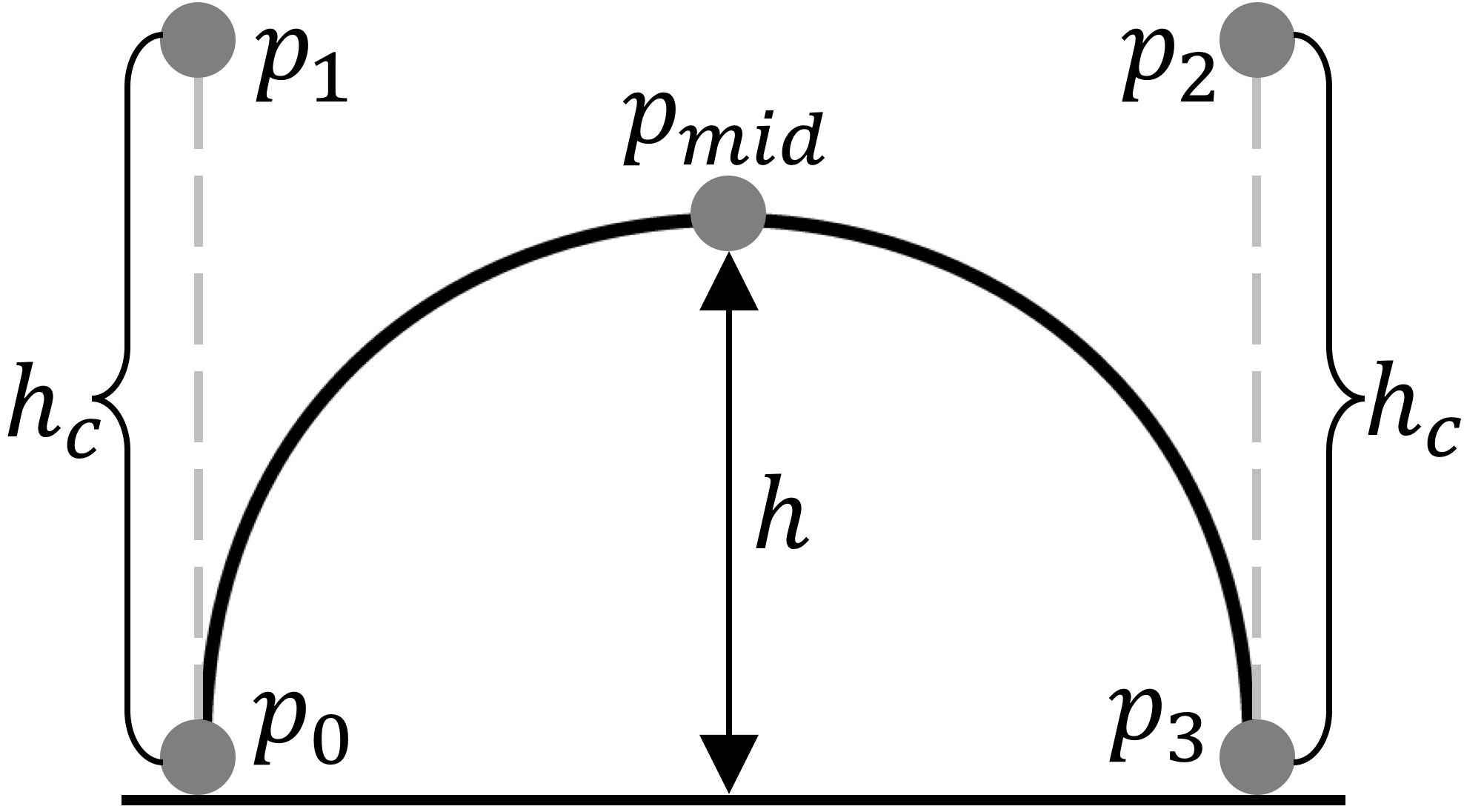}
\vspace{-1mm} 
\end{minipage}
of flow lines. $P_1$ and $P_2$ are set to the same height to ensure symmetry
such that the highest point will be at $t=0.5$, the mid-point of the tube. We can use the height of two control points ($h_c$) to precisely control the height of the mid-point ($h$):  $h_c = \frac{h}{6 \times 0.5^3} = \frac{4}{3}h$. 

Three different height encodings were evaluated: 

\noindent\textbf{\textit{3D constant:}} All flows have the same height (Fig. \ref{fig:first-study-vis}(c)).

\noindent\textbf{\textit{3D quantity:}} Height linearly proportional to flow quantity (Fig.\ \ref{fig:first-study-vis}(d)) such that small quantity flows will be at bottom, while large ones will be on top. In the pilot we also tried the inverse (smaller flows higher) but this was found to be severely cluttered.

\noindent\textbf{\textit{3D distance:}}  Height linearly proportional to Euclidean distance between the origin and destination (Fig. \ref{fig:first-study-vis}(e)). Close flows will be lower, while flows further apart will be on top. 
Again the inverse was tried in the pilot but quickly discarded as unhelpful.

\noindent\textbf{Encodings common to all conditions:} 
Quantity was encoded in all conditions using thickness of lines (in 2D) and diameter of tubes (in 3D). Several evaluations informed our choice of  direction encoding.  Holten \emph{et al.}\ evaluated encodings of unweighted edge direction in node-link diagrams~\cite{Holten:2009directed,Holten:2011fp} and found tapering of lines is the only direction encoding to be more effective than colour gradient. In the context of OD flow maps a study by Jenny \emph{et al.}~\cite{Jenny:2017ci} revealed difficulties interpreting tapered connections in geographic context. Furthermore, the  use of  line width/diameter to show weight makes tapered edges impractical (the only part of the line where width could be reliably compared would be at the origin).  We therefore chose to use colour gradient in both 2D and 3D conditions, using the same colour gradient (red-green) from~\cite{Holten:2011fp} to present direction information in our study.
To reduce the distortion of shapes and also for aesthetic reasons, we chose to use the Hammer map projection, an equal-area projection with an elliptical boundary.  \added{See \cite{Yang:2018mg} for additional details of our use of this projection.}
The Natural Earth raster map from \url{naturalearthdata.com} was used as the base texture. Originally, we had concerns about the texture colour interfering with flow readability but in our pilot tests participants had no problem with this. 
A legend was presented with all flow maps, indicating direction, quantity and  other encodings (e.g.\ height for distance).

\noindent\textbf{Rendering:}  
Geometry computation was accellerated with the GPU, tessellation was used for curve and tube interpolation, and a geometry shader was used to build the structure of line or tube segments.

\noindent\textbf{Interactions:} We provided the same interaction across the five visualisations. First, viewers can move in space to change their viewpoint. Second, we allowed viewers to change the 3D position and rotation of the map. They could pick up the map using a standard handheld VR controller, and reposition or rotate it in 3D space. We did not provide explicit interactive widgets or dedicated manipulations for adjusting the scale of the different maps. However, viewers could either move closer to the maps, or pick maps with a VR controller and bring them closer to their HMD. We did not allow other interaction such as filtering as we wished to focus on base-line readability of the representations.

\subsection{Experiment}

\noindent\textbf{Stimuli and Task Data:} We used  datasets based on real international migration flows between countries~\cite{Abel:2014iz} for the study. We show only a single net flow between each pair of countries. To control the number of flows for our different difficulty conditions we symmetrically filtered the data by dropping the same percentage of small and large flows. We randomised the origin and destination of the original dataset to ensure different data for each question.

Task: To keep the study duration for each participant to around one hour, we chose to evaluate a single task: finding and comparing the flow between two given pairs of locations:\\
\textit{For the two flows from A to B and X to Y, which is greater?}\\
Following Feiner \textit{et al.}~\cite{Feiner:1993ip}, leader lines were used to link  labels ``A'', ``B'' with the origin and destination of one flow and ``X'', ``Y'' with the origin and destination of the other flow. Labels were horizontal and rotated in real-time so as to remain oriented towards the viewer.

This task was chosen because it combines two fundamental sub-tasks: \emph{searching} for the flow line between two given locations and \emph{comparison} of magnitude of two flows. We would expect visual clutter to negatively impact on both of these sub-tasks while dual encoding of magnitude might help with comparison.
Besides the choice of flow representions, two factors may affect user performance: the number of flows, and the relative difference of quantity between two given flows.

Number of flows: In the study by Jenny \textit{et al.}~\cite{Jenny:2017ci}, the largest number of flows tested was around 40.  To better understand scalability, we decided to test three different difficulty levels in this study: (1) 40 flows with 20\% difference, (2) 40 flows with 10\% difference and (3) 80 flows with 20\% difference. We required the two flows in question to be separated by at least 15\textdegree~on the great circle connecting them so as to avoid situations where the origin and destination of a flow were too close to be clearly distinguished. To balance the difficulty of searching for a flow, all  origins and destinations of flows under comparison were required to have more than three flows.

Quantity encoding: The smallest (largest) flow magnitude was mapped to the thinnest (widest) flow width, and intermediate values were linearly encoded.

\noindent\textbf{Experimental Set-up:}
We used an HTC Vive with $110^{\circ}$ field of view and 90Hz refresh rate as the VR headset for the experiment. The PC was equipped with an Intel i7-6700K 4.0GHz processor and NVIDIA GeForce GTX 1080 graphics card. Only one handheld VR controller was needed in the experiment: participants could use this to reposition and rotate the map in 3D space. The frame rate was around 110FPS, \textit{i.e.} computation was faster than the display refresh rate.

\added{Visuals were positioned comfortably within the users' reach and sized by default to occupy approximately 60\% of the viewers' horizontal field of view.} The map was texture-mapped onto a quad measuring 1$\times$0.5 metre and placed at 0.55 metre in front and 0.3 metre under participants' eye position, and tilted to 45\textdegree. The map was centred on 0\textdegree~longitude and 0\textdegree~latitude. We repositioned the map at the beginning of every question. The thickness of lines in 2D and diameter of tubes in 3D were in the range of 2mm to 16mm. The height of \textit{3D quantity} and \textit{3D distance} was linearly mapped to the range of 5cm to 25cm. The constant height for \textit{3D constant} was 15cm.

\noindent\textbf{Participants:}
We recruited 20 participants (8 female) from our university.  All had normal or corrected-to-normal vision and included university students and researchers. 1 participant was under 20, 15 participants were within the age group 20$-$30, 1 participant was between 30$-$40, and 3 participants were over 40. VR experience varied: 13 participants had less than 5 hours of prior VR experience, 5 participants had 6$-$20 hours, and 3 participant had more than 20 hours.  \added{While our encoding used colour to indicate direction, the tasks used did not involve ambiguity regarding direction.  Therefore, we did not test participants for colour blindness.}

\noindent\textbf{Design and Procedure:}
The experiment was within-subjects: 20 participants $\times$ 5 visualisations $\times$ 1 task $\times$ 3 difficulty levels $\times$ 5 repetitions = 1,500 responses (75 responses per participant) with performance measures and lasted one hour on average. Latin square design was used to balance the order of visualisations. %

Participants were first given a brief introduction to the experiment. Before they put on the VR headset, we measured the pupil distance (PD) of the participants, and adjusted the PD on the VR headset. Two types of training were included in this experiment: interaction training and task training. Both were  conducted when each flow map representation was presented to the participants for the first time.

During \textit{interaction training} participants were introduced to the flow map with details of the encodings and given sufficient time to familiarise themselves with  interaction. They were then asked to pick up the map and put it on a virtual table in VR. This activity familiarised participants with each flow map representation as well as the VR headset and controller. This was followed by \textit{task training}. Two sample tasks
were given to participants with unlimited time. 
We asked the participants to check their strategies both when they were doing the training tasks and when the correct answers for those tasks were shown.

\begin{figure}[t!]
	\centering
	\includegraphics[width=0.5\textwidth]{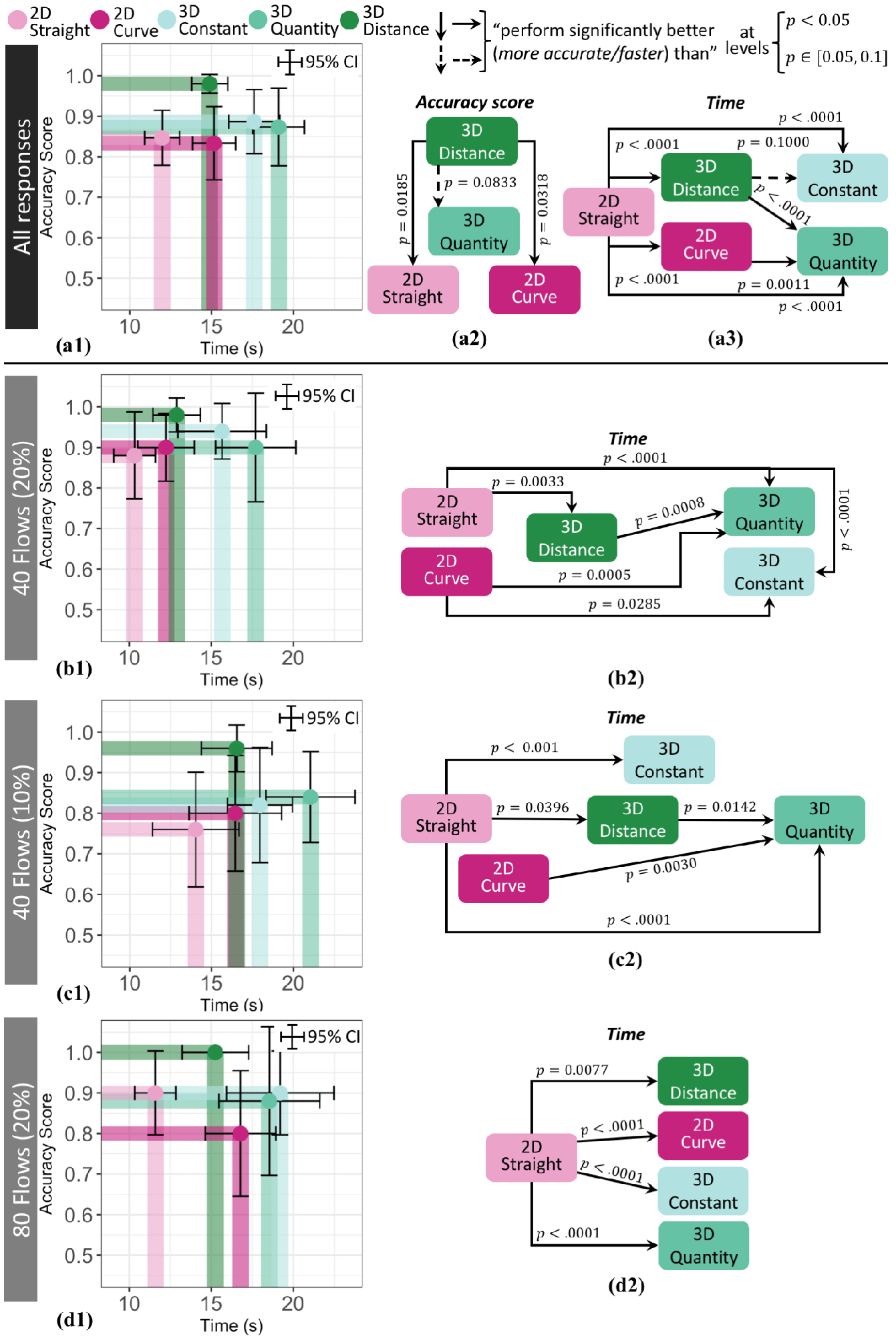}
	\caption{Study 1: Accuracy score and response time for different flow map representations in first study: (a1, b1, c1, d1) Average performance with 95\% confidence interval, (a2, a3, b2, c2, d2) graphical depiction of results of pairwise post-hoc test.}
	\label{fig:first-study-result-all}
\end{figure}

Participants were presented with the five flow map representations in counterbalanced order. A posthoc questionnaire recorded feedback on: (1) preference ranking of visualisations in terms of visual design and ease of use for the tasks, (2) advantages and disadvantages of each visualisation, (3) strategies for different flow maps, and (4) background information about the participant.
In the questionaire the visualisations were listed in the same order that they were presented to participants during the experiment.  \added{All experimental materials are available for download from \url{https://vis.yalongyang.com/VR-Flow-Maps.html}.}

\noindent\textbf{Measures:}
We measured the time between the first rendering of the visualisation and the double-click on the controller trigger button. After participants double-clicked, the visualisation was replaced by two buttons to answer the question. Collected answers were binary (i.e. participants chose between two options) and we therefore used the  accuracy score from \cite{Willett:2015fv} to indicate perfect performance with 1, and a result equal to pure chance (i.e. randomly guessing) with 0: $(\frac{number\ of\ correct\ responses}{number\ of\ total\ responses} - 0.5) \times2$.

We also recorded the number of clicks, head position, head rotation, controller position, and map position every 0.1s. Recording these parameters is important, as users can move in a relatively large open space with the HTC Vive HMD. 

\subsection{Results}
Accuracy scores were not normally distributed (checked with histograms and Q$-$Q plots). Significance was tested with the Friedman test because we have more than two conditions; the Wilcoxon-Nemenyi-McDonald-Thompson post-hoc test was used to compare pairwise \cite{Hollander:1999ns}.

Response times were log-normal distributed (checked with histograms and Q$-$Q plots), so a log-transform was used for statistical analysis \cite{howell2012statistical}.
We chose one-way repeated measures ANOVA with linear mixed-effects model to check for significance and applied Tukey's HSD post-hoc tests to conduct pairwise comparisons \cite{field2012discovering}.
For user preferences we again used the Friedman test and the Wilcoxon-Nemenyi-McDonald-Thompson post-hoc test to test for significance. 

The Friedman test revealed a statistically significant effect of visualisations on accuracy ($\chi^2(4) = 12.29, p = .015$). Fig.~\ref{fig:first-study-result-all}(a1) shows the average accuracy score of \textit{3D distance} (0.98) was higher than that of \textit{3D quantity} (0.87) and of 2D flow maps (\textit{straight} with 0.85 and \textit{curve} with 0.83). While \textit{3D distance} also outperformed \textit{3D constant} (0.89), this was not found to be statistically significant. A post-hoc test showed statistical significances as per Fig.~\ref{fig:first-study-result-all}(a2).

The ANOVA analysis showed significant effect of visualisations on time ($\chi^2(4) = 50.63, p < .0001$). \textit{2D straight} (avg.\ 12.0s) was significantly faster than other flow maps. \textit{3D distance} (avg.\ 14.9s) and \textit{2D curve} (avg.\ 15.2s) were significantly faster than \textit{3D constant} (avg.\ 17.6s) and \textit{3D quantity} (avg.\ 19.1s) (see Fig.~\ref{fig:first-study-result-all}(a3)).

By difficulty condition the Friedman test did not reveal significant effect for accuracy. The ANOVA analysis revealed:\\
\noindent\textbf{\textit{40 flows (20\%)}:} $\chi^2(4) = 36.39, p < .0001$. \textit{2D straight} (avg.\ 10.3s) was significantly faster than other flow maps except \textit{2D curve} (avg.\ 12.2s). \textit{3D distance} (avg.\ 12.9s) was only slower than \textit{2D straight}. \textit{3D quantity} (avg. 17.7s) was slower than other flow maps, except \textit{3D constant} (avg.\ 15.7s).\\
\noindent\textbf{\textit{40 flows (10\%)}:} $\chi^2(4) = 31.30, p < .0001$. \textit{2D straight} (avg.\ 14.0s) was significantly faster than other flow maps except \textit{2D curve} (avg.\ 16.4s). \textit{3D quantity} (avg.\ 21.0s) was significantly slower than \textit{3D distance} (avg. 16.5s). It also seemed to be slower than \textit{3D constant} (avg. 18.0s), but no statistical significance was found.\\
\noindent\textbf{\textit{80 flows (20\%)}:} $\chi^2(4) = 34.39, p < .0001$. \textit{2D straight} (avg.\ 11.6s) was significantly faster than other flow maps: \textit{2D curve} (avg.\ 16.8s), \textit{3D distance} (avg.\ 15.3s), \textit{3D constant} (avg.\ 19.2s) and \textit{3D quantity} (avg.\ 18.5s).

\begin{figure}
	\centering
	\includegraphics[width=0.44\textwidth]{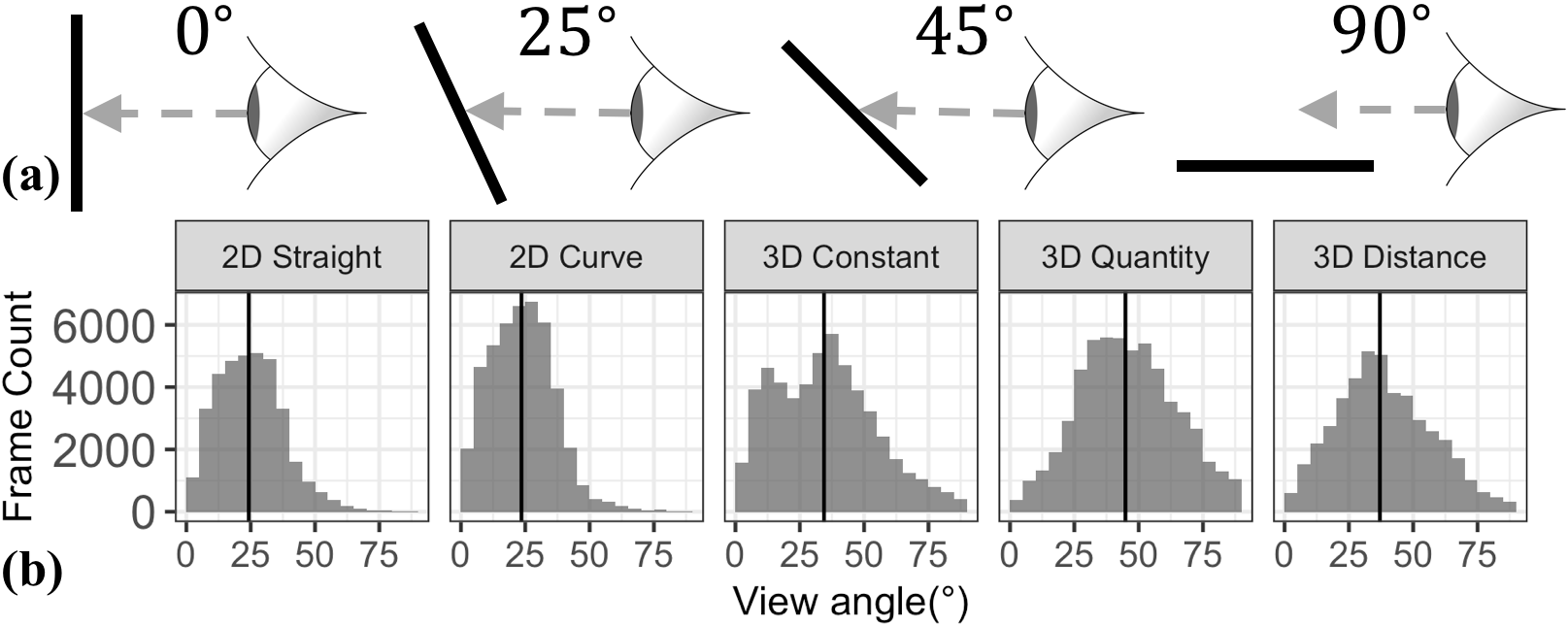}
	\caption{(a)~Demonstration of different view angles, (b)~view angle distribution among different flow maps with median line.}
	\label{fig:first-angle-distribution}
\end{figure}
\noindent\begin{minipage}{0.55\columnwidth}
 When analysing the details of interaction, we sampled every second frame. If the head or the map moved more than 1cm or rotated more than 5\textdegree, we considered it as an interaction, and accumulated the interaction time for every user and then normalised the time related to the percentage of time spent on that question. Friedman test revealed a
 \end{minipage}
\begin{minipage}{3.8cm}
\vspace{-4mm}
\centering
\includegraphics[width=4cm]{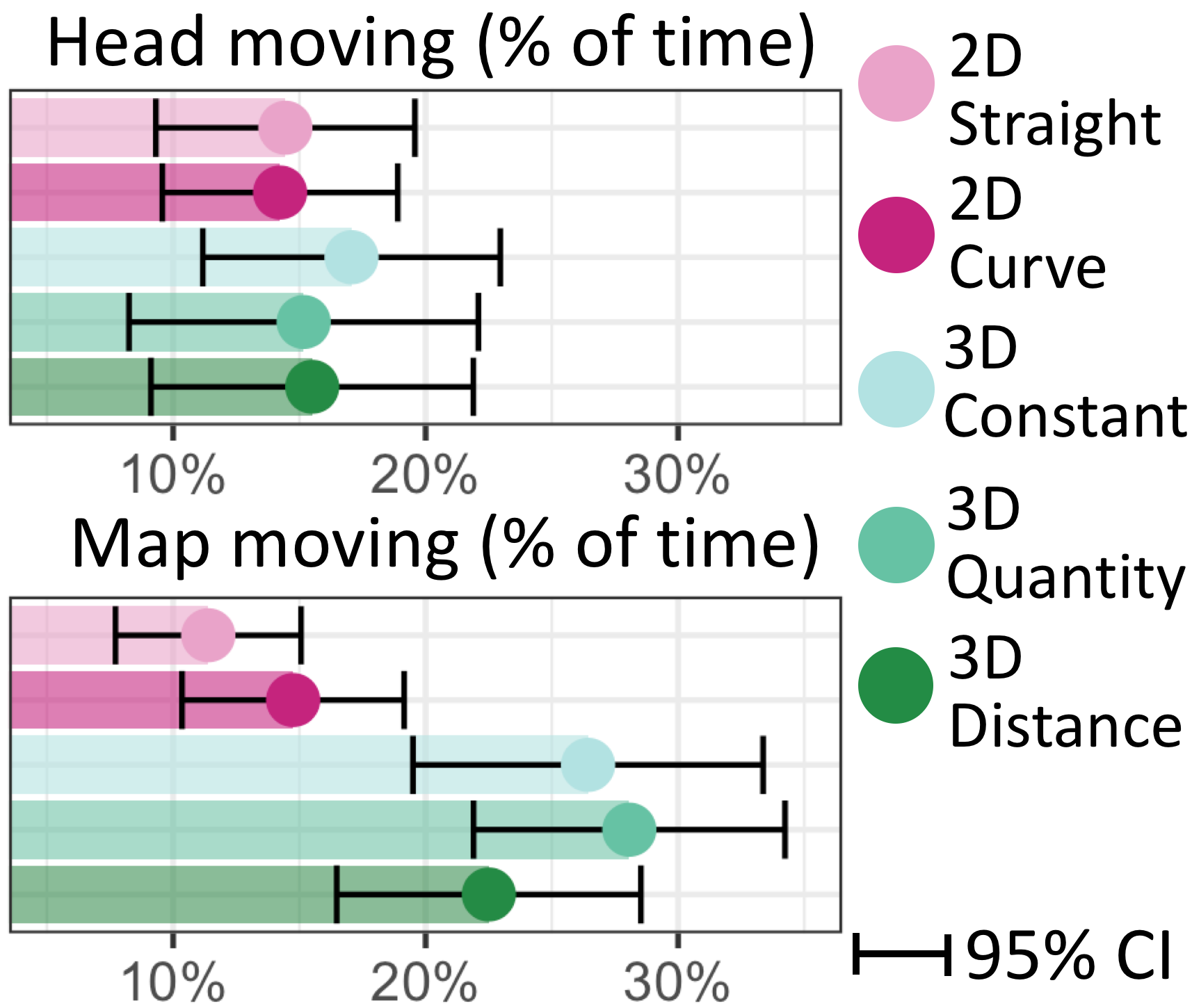} 
\vspace{-4mm}
\end{minipage}  
statistically significant effect of visualisations on map movements ($\chi^2(4) = 30.4, p < .0001$). Participants tended to move the map significantly more in all 3D conditions than 2D conditions (all $p < .05$). Wilcoxon signed rank test also revealed participants spent statistically significant more percentage of time moving the map than their head in \textit{3D distance} (at level $p=.10$) and \textit{3D quantity} (at level $p=.06$).

We also analyzed the view angle, i.e. the angle between viewers' heads forward vector and the normal vector of the map plane.  The Friedman test revealed that the effect of visualisations on the percentage of time spent with a view angle larger than 45\textdegree~per user  was statistically significant  ($\chi^2(4) = 64.88, p < .0001$). As one might expect, participants spent significantly more percentage of time with large view angles in all 3D conditions than all 2D conditions (see Fig.~\ref{fig:first-angle-distribution}). 

\noindent\textbf{User preference and feedback:}\\
\noindent\begin{minipage}{0.49\columnwidth}
Participant ranking for each of the four visualisations by percentage of respondents is shown by colour: \setlength{\fboxsep}{1.5pt}\colorbox{first}{\textcolor{white}{$1^{st}$}}, \setlength{\fboxsep}{1pt}\colorbox{second}{\textcolor{white}{$2^{nd}$}}, \setlength{\fboxsep}{1pt}\colorbox{third}{$3^{rd}$}, \setlength{\fboxsep}{1pt}\colorbox{fourth}{$4^{th}$} and \setlength{\fboxsep}{1pt}\colorbox{fifth}{$5^{th}$}. 
For \emph{visual design}, the Friedman test revealed a significant effect of visualisations on preference ($\chi^2(4)$
\end{minipage}
\begin{minipage}{4.4cm}
\centering
\vspace{0mm}
\small
\fontfamily{phv}\selectfont
Visual Design Ranking\\
\includegraphics[width=4.6cm]{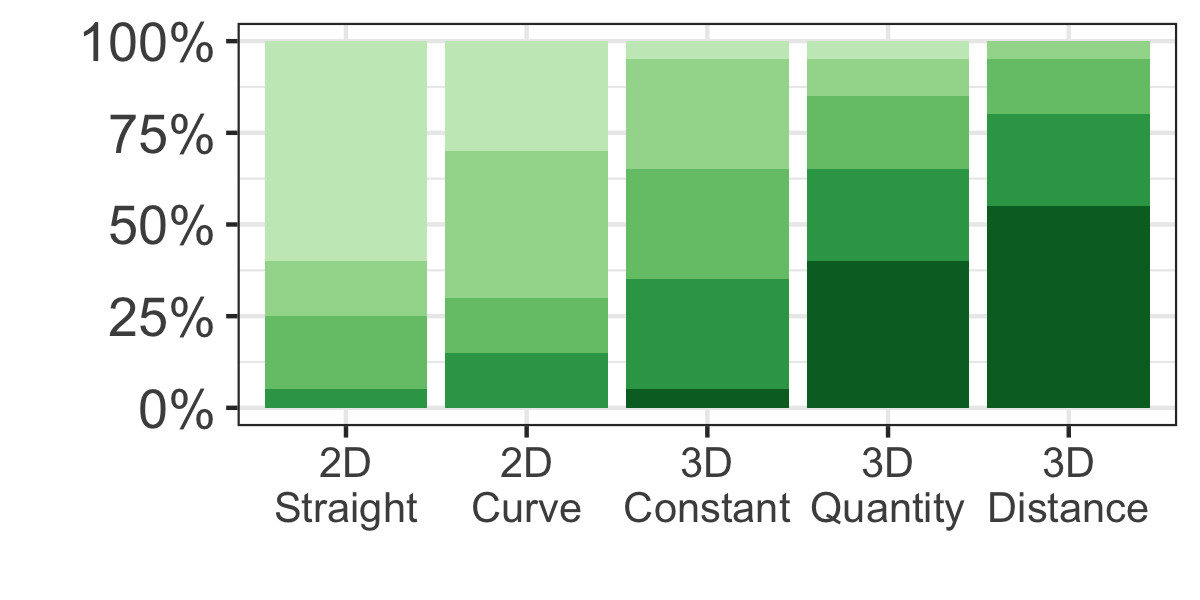} 
\vspace{-4mm}
\end{minipage} 
$=38.6, p < .0001$). The strongest preference was for \textit{3D distance}, with 95\% voting it as top three. The post-hoc tests also found a stronger preference for \textit{3D distance} than \textit{3D constant} (65\% voting it as top three), \textit{2D curve} (40\% voting it as top three) and \textit{2D straight} (30\% voting it as top three). Participants also seemed to prefer \textit{3D quantity} (85\% voting it as top three), however, the post-hoc tests only suggested it was preferred to \textit{2D straight}. 

\noindent\begin{minipage}{0.49\columnwidth}
For \emph{readability}, the Friedman test indicated significant effect of visualisations on preference ($\chi^2(4)=23.32, p = .0001$). The strongest preference is again for the \textit{3D distance}, with 90\% of respondents voting it top three. The post-hoc
\end{minipage}
\begin{minipage}{4.4cm}
\centering
\vspace{0mm}
\small
\fontfamily{phv}\selectfont 
\hspace{-3mm}Readability Ranking\\
\vspace{0mm}
\includegraphics[width=4.6cm]{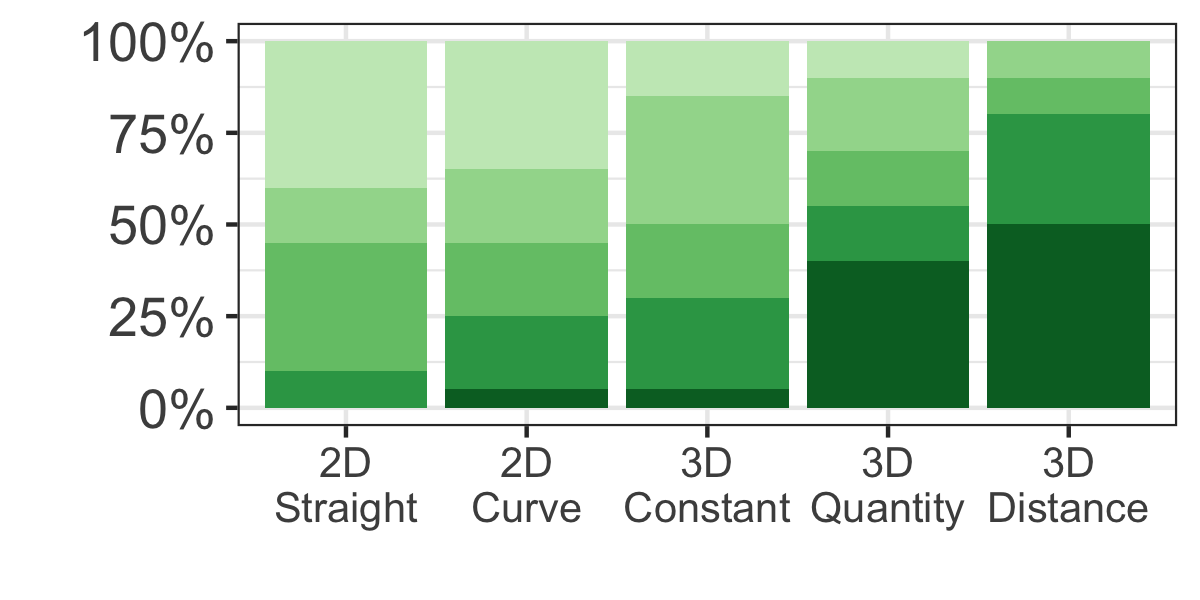} 
\vspace{-4mm}
\end{minipage} 
tests again showed stronger preference for \textit{3D distance} than \textit{3D constant} (50\% voting it as top three), \textit{2D curve} (45\% voting it as top three) and \textit{2D straight} (45\% voting it as top three). Participants also seemed to prefer \textit{3D quantity} (70\% voting it as top three), however, the post-hoc tests again only revealed it was preferred over \textit{2D straight}. 

The final section of the study allowed participants to give feedback on the pros and cons of each design. Qualitative analysis of these comments reveal (overall):

\noindent\textbf{\textit{2D straight}} was found to be easy for small data sets, however, lines were found to be hard to distinguish due to increasing overlaps in large data sets. Several participants reported: \emph{``I answered sometimes with a very low confidence, close to luck.''}

\noindent\textbf{\textit{2D curve}} was found to have fewer overlaps than \textit{2D straight}. However, many participants reported the curvature made it difficult to follow lines, and sometimes, the curvature was found to be unexpected, which apparently increased difficulty.

\noindent\textbf{\textit{3D constant}} was found to have considerable numbers of overlaps by most participants. However, some participants also found it efficient with interaction: \emph{``I could look at the line from a straight angle, plus wiggle the map around a little to confirm the line.''} 

\noindent\textbf{\textit{3D quantity}} was more trusted. Many participants reported: \emph{``If I couldn't work out from thickness, I could move the visualisation to compare heights from the side to confirm my answer. I felt more confident.''} However, they also commented about the extra time they spent for this confirmation.

\noindent\textbf{\textit{3D distance}} was easy to distinguish flows. \emph{``This one left enough gaps between the curves to clearly distinguish the curves''} and \emph{``it was visually appealing.''} However, a few participants commented they felt more confident with \textit{3D quantity}, and a few commented that this might be due to the double encoding used by \textit{3D quantity} (\textit{3D quantity} encodes quantity with height and width).

\subsection{Key Findings}
The main finding of this study was that the \textit{3D distance} was more accurate than the other 3D conditions and both 2D visualisations. It was also the preferred visualisation. We also found that:
\begin{itemize}[leftmargin=1em]
	\item The \textit{2D straight-line} flow map was the fastest in almost all conditions but least preferred.	
	 \item Participants tended to look more often from the side in 3D conditions than in 2D conditions.
	\item Participants tended to interact with the map more in 3D conditions than 2D conditions.
	\item Participants tended to move the map more than their heads in \textit{3D distance} and \textit{3D quantity}.
\end{itemize}

\hyphenation{Maps-Link}
\section{Study 2: Flows on Flat Maps, Globes and Map Pairs}
\label{sec:study2}
In the second study, we focused on exploring different 2D and 3D representations of the reference space.

\subsection{Visualisations and Interactions}
\label{sec:globes-mapslink}
We evaluated 4 different representations. 

\noindent\textbf{\textit{2D straight}} and \textbf{\textit{3D distance}}: The first two used a flat map to represent the reerence space. These were the best performing representations from our first study (fastest with  \textit{\textit{2D straight}}, most accurate with \textit{\textit{3D distance}}). 

\begin{figure}
	\centering
	\includegraphics[width=0.42\textwidth]{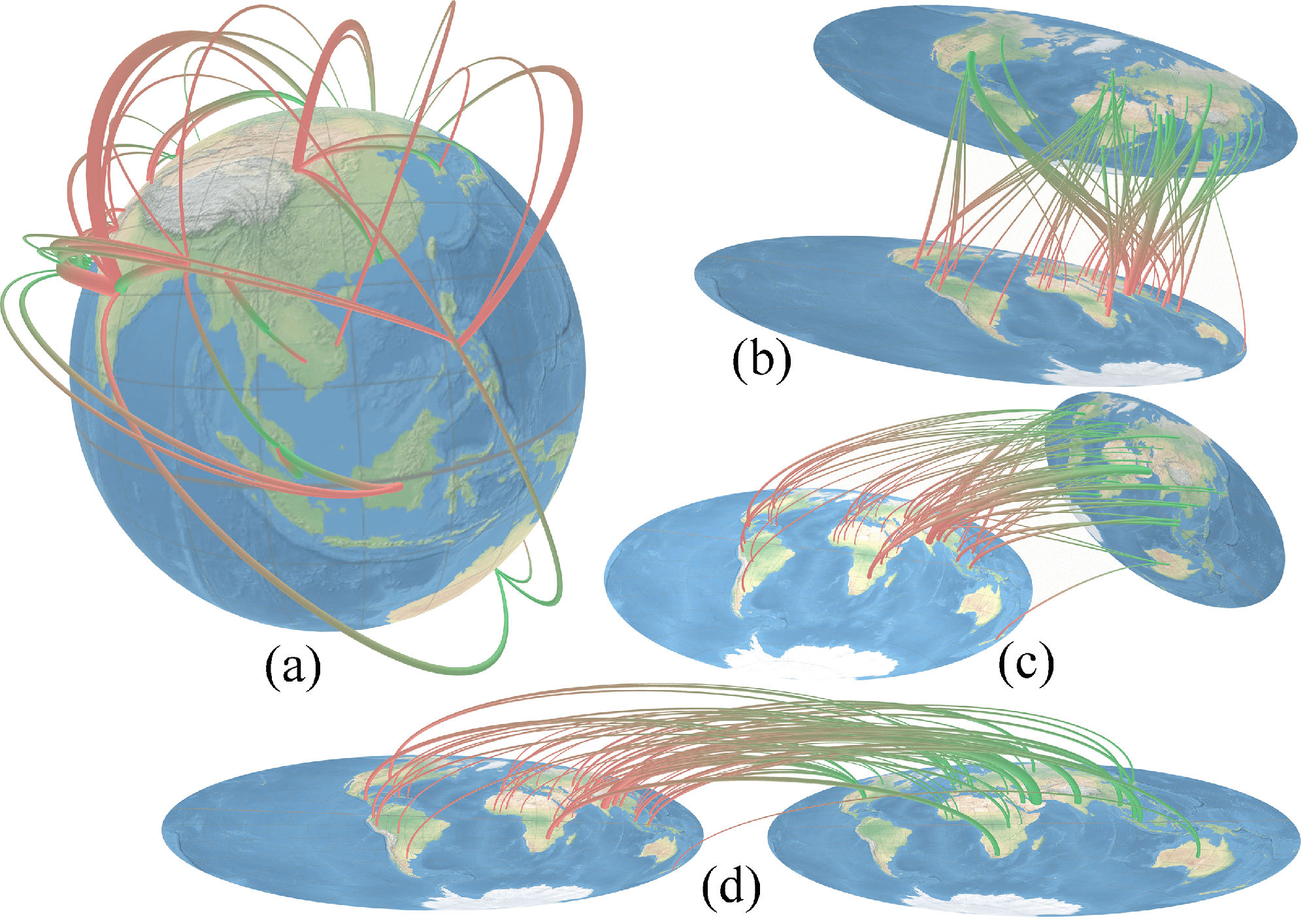}
	\caption{Study 2: (a) 3D globe flow map, (b, c, d) MapsLink: flow tubes linking a pair of flat maps.}
	\label{fig:second-study-vis}
\end{figure}

\noindent\textbf{\textit{Globe}:} A 3D globe has proven to be an effective way to present global geometry \cite{Yang:2018mg} but the effectiveness for showing OD flows has not been previously evaluated. 
We represented flow in the globe using 3D tubes, i.e. we linked two locations on the globe with their great circle trajectory (Fig. \ref{fig:second-study-vis}(a)).
Based on the result of our first study, we chose to use curve height to encode the great circle distance between two points where height here refers to the distance between the curve's centre point and the centre of the globe. 
We used a cubic transformation with interpolation factor $0 \le t \le 1$, $h_t = ((-|t-0.5| / 0.5)^3 + 1) \times h + radius$.
 
\noindent\textbf{\textit{MapsLink}:} We also evaluated a novel flow map representation which used a separate reference space for the origin and destination. 
This used two flat maps in 3D space: the \textit{origin map} showing  origins and the \textit{destination map} showing  destinations (Fig. \ref{fig:second-study-vis}~(b)). Flows from origin to destination were rendered with curved tubes linking origins in the origin map and destinations in the destination map. The 3D tubes were cubic B\'ezier curves (see Equ. \ref{equ:bezier}) with orgin and destination as the first and last control point. As the two maps might not be in the same plane, we could not control the height in the same way as the first experiment, instead, the two control points were raised from the orign and destination maps to the same height, which was proportional to the Euclidean distance in 3D space between the origin and destination points. This meant when the origin and destination map were  facing each other, origins and destinations were linked by straight lines and by smooth curves at other orientations. 

\label{sec:extra-interaction} 
\noindent{\textbf{Interactions:}}
As in the first study participants could rotate and reposition the visualisation. In addition we allowed participants to adjust the centre of the geographic area in the visualisation in VR. Viewers could pick any location and drag it to a new position using the VR controller. This interaction (called \textit{geo-rotation}) was presented in~\cite{Yang:2018mg}
and allows the viewer \added{to} bring the geographic area of interest to the centre of the visualisation. This changes the relative position of points and straight line flows on the flat map, thus providing some of the benefits that changing viewpoint provided for the 3D representations of flow. We were interested to see if it improved accuracy of the flat map with straight line 2D flows. 

\subsection{Experiment}
\noindent\textbf{Stimuli and tasks:}
The same task of finding and comparing flows between two origin-destination pairs was used in this study. The same raw data was used as well. With the addition of the \textit{geo-rotation} interaction, we assumed participants could complete the task more easily. We therefore increased the difficulty of the three conditions: (a) 80 flows with 20\% difference, (b) 80 flows with 10\% difference and (c) 120 flows with 20\% difference. Pilots demonstrated participants could handle these difficulty conditions.

\noindent\textbf{Set-up:}
The headset and PC setup were the same as used in Study 1, except that two controllers were given to the participants so that they could use one controller to position and rotate the map/globe while the other is used for \textit{geo-rotation}.
In \textit{MapsLink} two controllers also affords bimanual gestures to manipulate both maps simultaneously.

The \textit{2D straight} and \textit{3D distance} setup was the same as for Study 1. As for was the case for 3D maps in Study 1, the thickness of tubes in \textit{globe} and \textit{MapsLink} was linearly mapped to the range of 0.1cm to 0.8cm.

The globe had a radius of 0.4 metre. The starting position for the centre of the globe was 1 metre in front and 0.3 metre under the participant's eye position. The geographic centre of the globe was set at 0\textdegree longitude and 0\textdegree latitude, facing towards the viewer. As for \textit{3D distance}, the height was linearly mapped to the range of 5cm to 25cm.

The two maps of \textit{MapsLink} measure 75\% of the size of the flat map in the first study (0.75 $\times$ 0.375 metre). We reduced size to reduce the chance of the two maps intersecting. The two maps were first placed 0.55 metre in front and 0.3 metre under participant's eye position, then the origin map was moved left 0.4 metre, and the destination map was moved right 0.4 metre. Finally, both maps were tilted towards participants around the y-axis by 30\textdegree and around the x-axis by 45\textdegree. Flows were modeled with cubic B\'ezier curves: The two control points were placed on a line orthogonal to map planes; distances between control points and planes were between 5 and 50cm, and were proportional to the distances between origins and detinations (which were assumed to be between 0 and 2m).

\noindent\textbf{Participants:}
We recruited 20 participants (6 female) from our university campus, all with normal or corrected-to-normal vision. Participants included university students and researchers. 14 participants were within the age group 20$-$30, 5 participants were between 30$-$40, and 1 participant was over 40. VR experience varied: 14 participants had less than 5 hours of prior VR experience, 4 participants had 6$-$20 hours, and 2 participants had more than 20 hours.

\noindent\textbf{Design and Procedure:}
A similar design to the first user study was used, within-subjects: 20 participants $\times$ 4 visualisations $\times$ 1 task $\times$ 3 difficulty levels $\times$ 5 repetitions = 1,200 responses (60 responses per participant) with performance measures and lasted one hour on average. Latin square design was used to balance the order of visualisations, and 4 data sets were ordered  to balance the effect of tasks across participants (i.e. every flow map was tested on all data sets).

\begin{figure}[t!]
	\centering
	\includegraphics[width=0.48\textwidth]{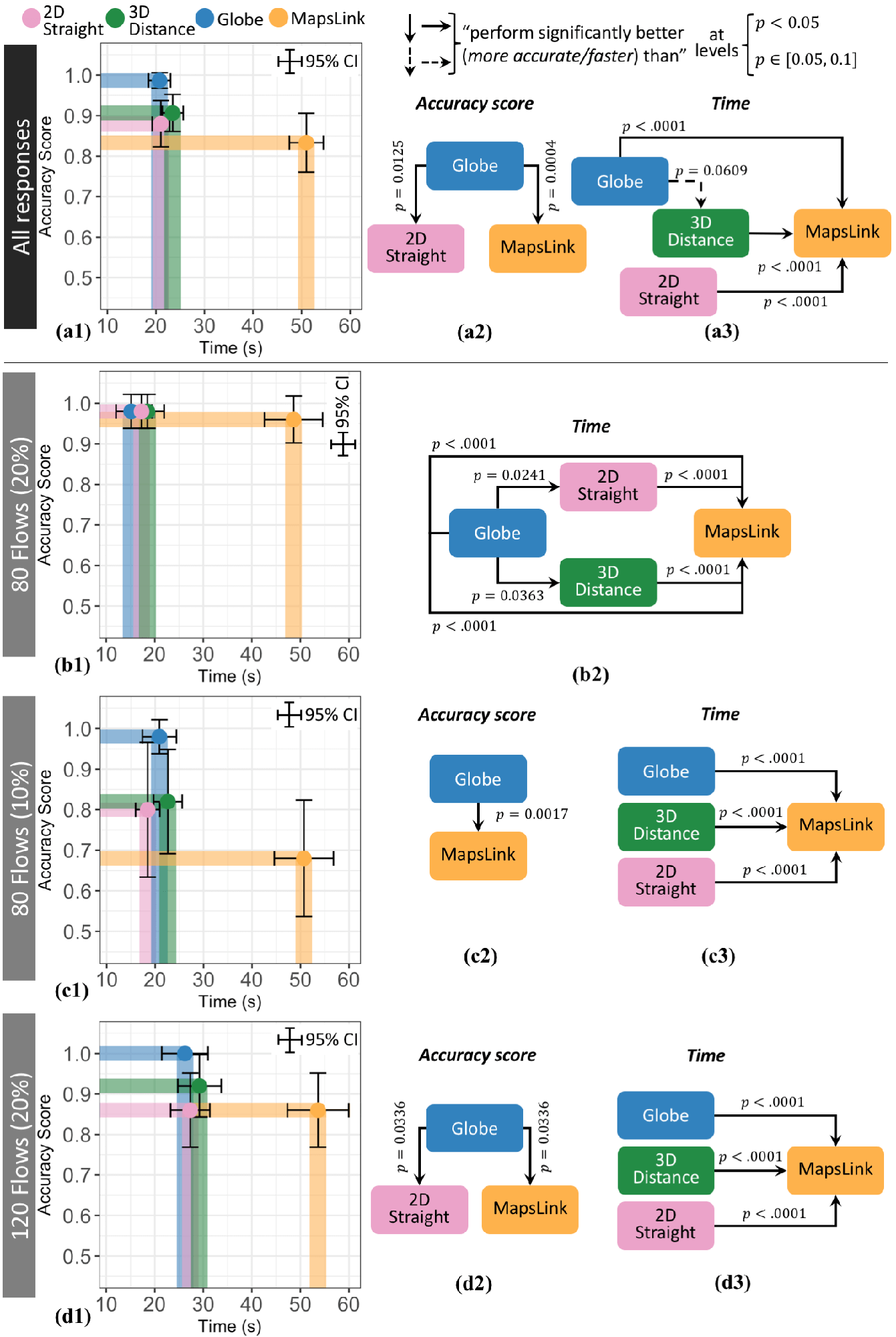}
	\caption{Study 2: (a1, b1, c1, d1) Average performance with 95\% confidence interval, (a2, a3, b2, c2, c3, d2, d3) graphical depiction of results of pairwise post-hoc test.}
	\label{fig:second-study-result-all}
\end{figure}

The procedure was similar to Study 1 but with two modifications.
In \textit{interaction training}, in addition to asking participants to place the flow maps on top of a table, we also asked them to use \textit{geo-rotation} to rotate Melbourne to the centre of the map or to the centre of participant's view. In the \textit{posthoc questionnaire}, we added a question to rate their confidence with each flow map with a five-point-Likert scale.

\noindent\textbf{Measures:}
In addition to real-time recording of participant's head, controller and map  position and rotation information, we also recorded the time duration whenever a participant used \textit{geo-rotation}.

\subsection{Results}
As in the first study, after checking normality of the data with histograms and Q$-$Q, we used the Friedman test to check for significance of accuracy score and applied the Wilcoxon-Nemenyi-McDonald-Thompson post-hoc test to conduct pairwise comparisons. For response time, we chose one-way repeated measures ANOVA with linear mixed-effects model to check for significance of its $log$ transformed values and applied Tukey's HSD post-hoc to conduct pairwise comparisons. 

The Friedman test revealed a statistically significant effect of visualisations on accuracy ($\chi^2(3) = 18.06, p = 0.0004$). Fig.~\ref{fig:second-study-result-all}~(a1) shows the average accuracy score of \textit{globe} (avg. 0.99) was higher than that of \textit{2D straight} (avg. 0.88) and of \textit{MapsLink} (avg. 0.83). While \textit{globe} also outperformed \textit{3D distance} (avg. 0.91), this was not found to be statistically significant. A post-hoc test showed statistical significances as per Fig.~\ref{fig:second-study-result-all}~(a2).

The ANOVA analysis showed significant effect of visualisations on time ($\chi^2(3) = 107.87, p < .0001$). \textit{MapsLink} (avg. 50.9s) was significantly slower than other visualisations. \textit{Globe} (avg. 20.7s) was significantly faster than \textit{3D distance} (avg. 23.4s). \textit{2D straight} (21.0s) had no significances between \textit{globe} and \textit{3D distance}.

By difficulty condition the Friedman test revealed a significant effect on accuracy score:\\
\noindent\textbf{\textit{80 flows (20\%)}:} $\chi^2(3) = 0.69, p = .8750$. All visualisations had similar performance in this condition.\\
\noindent\textbf{\textit{80 flows (10\%)}:} $\chi^2(3) = 13.28, p = .0041$. \textit{Globe} (avg. 0.98) was significantly more accurate than \textit{MapsLink} (avg. 0.68). \textit{2D straight} (avg. 0.80) and \textit{3D distance} (avg. 0.82) had no statistical significance with other visualisations.\\
\noindent\textbf{\textit{120 flows (20\%)}:} $\chi^2(3) = 9.9, p = .0194$. Responses with \textit{globe} were perfect (with an accuracy score 1), and it was significantly more accurate than both \textit{2D straight} (avg. 0.86) and \textit{MapsLink} (avg. 0.86). \textit{3D distance} (avg. 0.92) had no statistical significance with other visualisations.

By difficulty condition the ANOVA analysis revealed significant effect on time:\\
\noindent\textbf{\textit{80 flows (20\%)}:} $\chi^2(3) = 97.62, p < .0001$. \textit{MapsLink} (avg. 48.6s) was significantly slower than other visualisations. \textit{Globe} (avg. 15.1s) was also significantly faster than \textit{2D straight} (avg. 17.3s) and \textit{3D distance} (avg. 18.5s).\\
\noindent\textbf{\textit{80 flows (10\%)}:} $\chi^2(3) = 71.6, p < .0001$. Again, \textit{MapsLink} (avg. 50.7s) was significantly slower than other visualisations: \textit{2D straight} (avg. 18.5s), \textit{3D distance} (avg. 22.6s) and \textit{globe} (avg. 20.9s).\\
\noindent\textbf{\textit{120 flows (20\%)}:} $\chi^2(3) = 58.72, p < .0001$. Again, \textit{MapsLink} (avg. 53.6s) was found to be significantly slower than other visualisations: \textit{2D straight} (avg. 27.3s), \textit{3D distance} (avg. 29.2s) and \textit{globe} (avg. 26.2s).

\begin{figure}
\setlength{\belowcaptionskip}{0cm}
	\centering
	\includegraphics[width=0.5\textwidth]{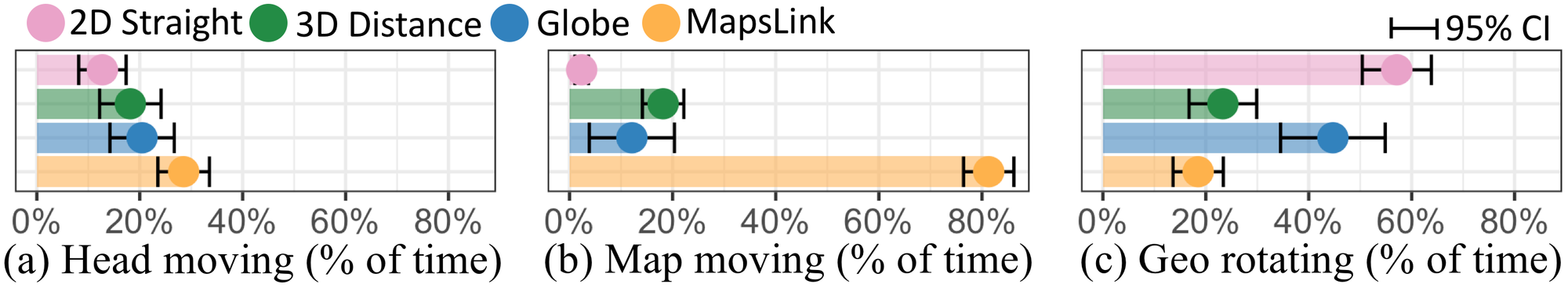}
	\caption{Study 2: Interaction time percentage with 95\% confidence interval}
	\label{fig:second-interaction-percentage}
\end{figure}

\noindent\textbf{Interactions:}
The percentage of time spent in different interactions per user was investigated (see Fig.~\ref{fig:second-interaction-percentage}). The Friedman test was used to determine statistical significance between different visualisations and between different interactions.

For \textit{head movement}, participants tended to spend a smaller percentage of time moving their heads in \textit{2D straight} than \textit{3D distance} ($p=.09$), \textit{globe} ($p=.03$) and \textit{MapsLink} ($p<.0001$). In \textit{map movement}, participants tended to move \textit{MapsLink} significantly more than other visualisations and there was more head movement for \textit{3D distance}  than for \textit{2D straight} (all $p<.05$). In \textit{geo-rotation}, participants used \textit{geo-rotation} significantly more in \textit{globe} and in \textit{2D straight} than with \textit{3D distance} and \textit{MapsLink} (all $p<.05$).

In \textit{2D straight}, the percentage of time was significantly different across interaction types: \textit{geo-rotation} $>$ head movement $>$ map movement (all $p<.05$). In \textit{3D distance}, no significant difference was found among different interactions. In \textit{globe}, \textit{geo-rotation} $>$ map movement ($p=.0025$). In \textit{MapsLink}, map movement $>$ head movement $>$ \textit{geo-rotation} (all $p<.05$).

We also investigated the benefits of adding \textit{geo-rotation} to \textit{2D straight} and \textit{3D distance}. We compared the responses of \textit{80 flows, 20\%} in the first (without \textit{geo-rotation}) and second (with \textit{geo-rotation}) studies.

For \textit{2D straight}, Exact Wilcoxon-Mann-Whitney test revealed \cite{Hothorn:2008fy} an increase of accuracy with \textit{geo-rotation} at level $p = .0530$ with $Z=-1.5072$. Log-transformed time values have been analysed with mixed ANOVA, the result demonstrated a significant increase in response time with \textit{geo-rotation} ($\chi^2(1) = 13.07, p < .0001$). For \textit{3D distance}, Exact Wilcoxon-Mann-Whitney test and mixed ANOVA did not show a significant difference between with and without \textit{geo-rotation}. 

\begin{figure}
	\centering
	\includegraphics[width=0.45\textwidth]{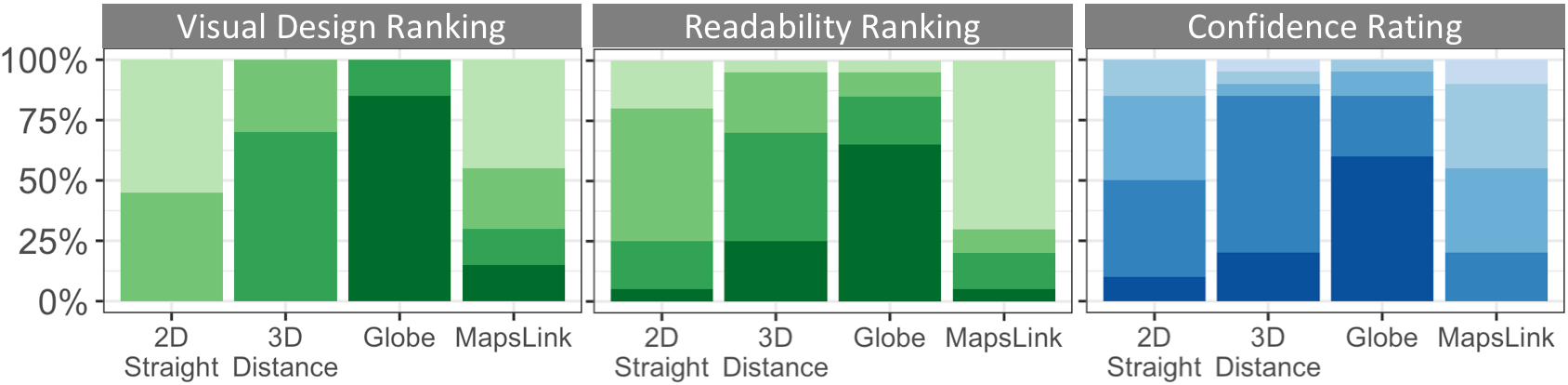}
	\caption{Study 2: Participants preference ranking (\setlength{\fboxsep}{1.5pt}\colorbox{first}{\textcolor{white}{$1^{st}$}}, \setlength{\fboxsep}{1pt}\colorbox{second}{\textcolor{white}{$2^{nd}$}}, \setlength{\fboxsep}{1pt}\colorbox{third}{$3^{rd}$} and \setlength{\fboxsep}{1pt}\colorbox{second_fourth}{$4^{th}$}) and confidence rating (from \setlength{\fboxsep}{1pt}\colorbox{first_conf}{\strut \textcolor{white}{fully confident}} to \setlength{\fboxsep}{2pt}\colorbox{last_conf}{not confident at all}).}
	\label{fig:second-rankings}
\end{figure}

\noindent\textbf{User preference and feedback:}
Participant ranking for each of the four visualisations by percentage of respondents is shown by colour (see Fig. \ref{fig:second-rankings}). For \emph{visual design}, the Friedman test revealed a significant effect of visualisations on preference ($\chi^2(3)=38.58, p < .0001$). The strongest preference was for the \textit{globe}, with 100\% voting it in the top two. The post-hoc tests also proved the strongest preference for \textit{globe} compared to other flow maps with all $p < .05$. \textit{3D distance} (70\% voting it top two) was also statistically preferred to \textit{2D straight} (0\% voting it top two). \textit{MapsLink}, with 30\% voting it top two, did not show statistical difference between \textit{2D straight} or \textit{3D distance}. For \emph{readability}, the Friedman test indicated significant effect of visualisations on preference ($\chi^2(3)=25.5, p < .0001$). The strongest preference is again for the \textit{globe}, with 85\% of respondents voting it top two. The post-hoc revealed stronger preference of \textit{globe} than \textit{2D straight} (25\% voting it top two) and \textit{MapsLink} (20\% voting it as top two). \textit{3D distance} (70\% voting it top two) was also statistically preferred to \textit{MapsLink}.

A five-point-Likert scale was used for rating participants' confidence (see Fig. \ref{fig:second-rankings}). The Friedman test revealed a significant effect of visualisations on confidence ($\chi^2(3)=24.82, p < .0001$). Participants felt significantly more confident in \textit{globe} and \textit{3D distance} than \textit{MapsLink}. \textit{Globe} was also found more confident than \textit{2D straight} (all $p<.05$).

The final section of the study allowed participants to give feedback on the pros and cons of each design. Qualitative analysis of these comments reveal (overall):

\noindent\textbf{\textit{2D straight}} was found to be very difficult at the beginning. However, many participants commented: \emph{``With the map rotating, it is usually possible to keep track of the pair of points.''}

\noindent\textbf{\textit{3D distance}} was found to be more visually appealing than \textit{2D straight}, and more efficient than \textit{2D straight} for small data. However, \emph{``things become very difficult when data size increases''}, and \emph{``sometimes, it felt more difficult than 2D (straight) map''}.

\noindent\textbf{\textit{Globe}} was found to be the most intuitive. Many participants also commented that the visualisation \emph{``felt very sparse so it was easy to tell which lines were connected to the points''}. Some participants also suggested to have a snapshot functionality to store the current globe rotation or two globes  positioned side by side.

\noindent\textbf{\textit{MapsLink}}: was found to be \emph{``very interesting to play with, but very difficult to use when it comes to the questions''}. However, some participants liked the freedom of manipulating it: \emph{``You can almost find the certain answer for each question by patiently manipulating map positions and rotating the maps.''} Meanwhile, many participants reported it took a long time to answer questions.

\subsection{Key Findings}
The main finding of this study was that user performance with the \textit{globe} was significantly more accurate than with \textit{2D straight} and \textit{MapsLink}. There was also some evidence that the \textit{globe} was more accurate than \textit{3D distance}, but this was not statistically significant. There was also some evidence that the \textit{globe} is resistant to increased clutter density (performance was stable with increasing flows, while other visualisations degraded with the number of flows). Additionally, we found:
\begin{itemize}[leftmargin=1em]
	\item \textit{MapsLink} was significantly slower than other representations and that participants spent most of their time moving the maps in \textit{MapsLink} (more than 80\% in average). 
	\item \textit{Geo-rotation} increased accuracy and slowed response time for \textit{2D straight}. With \textit{geo-rotation} there was no longer a significant difference in accuracy or speed between \textit{2D straight} and \textit{3D distance}.
	\item \textit{Globe} had the strongest preference in terms of visual design, while participants were more confident with both \textit{globe} and \textit{3D distance} and preferred them for readability.
	\item Participants chose to use different interactions in different representations. Compared to other visualisations, participants do not like to move their heads in \textit{2D straight} and participants liked to use \textit{geo-rotation} in \textit{2D straight} and \textit{globe}.
\end{itemize}

\hyphenation{Maps-Link}
\section{Study 3: Dense Flow Data Sets}
\label{sec:study3}
The third study was designed to investigate the scalability of the different flow maps. As participants spent significantly more time on \textit{MapsLink} and qualitative feedback indicated limited scalability for this design, we decide to test only the other three visualisations: \textit{2D straight}, \textit{3D distance} and \textit{globe}. We tested them with 200 and 300 flows, both with 10\% difference. 

We recruited 12 participants (6 female) from our university campus, all with normal or corrected-to-normal vision. Participants included university students and researchers. 9 participants were within the age group 20--30, 2 participant was between 30--40, and 1 participant was over 40. VR experience varied: 10 participants had less than 5 hours of prior VR experience, 2 participants had 6--20 hours.

Otherwise, experimental design and setup was identical to Study 2, within-subjects: 12 participants $\times$ 3 visualisations $\times$ 1 task $\times$ 2 difficulty levels $\times$ 8 repetitions = 576 responses (48 responses per participant) with performance measures and duration of one hour on average. 

\subsection{Results} 
\begin{figure}
\setlength{\belowcaptionskip}{-0.1cm}
	\centering
	\includegraphics[width=0.45\textwidth]{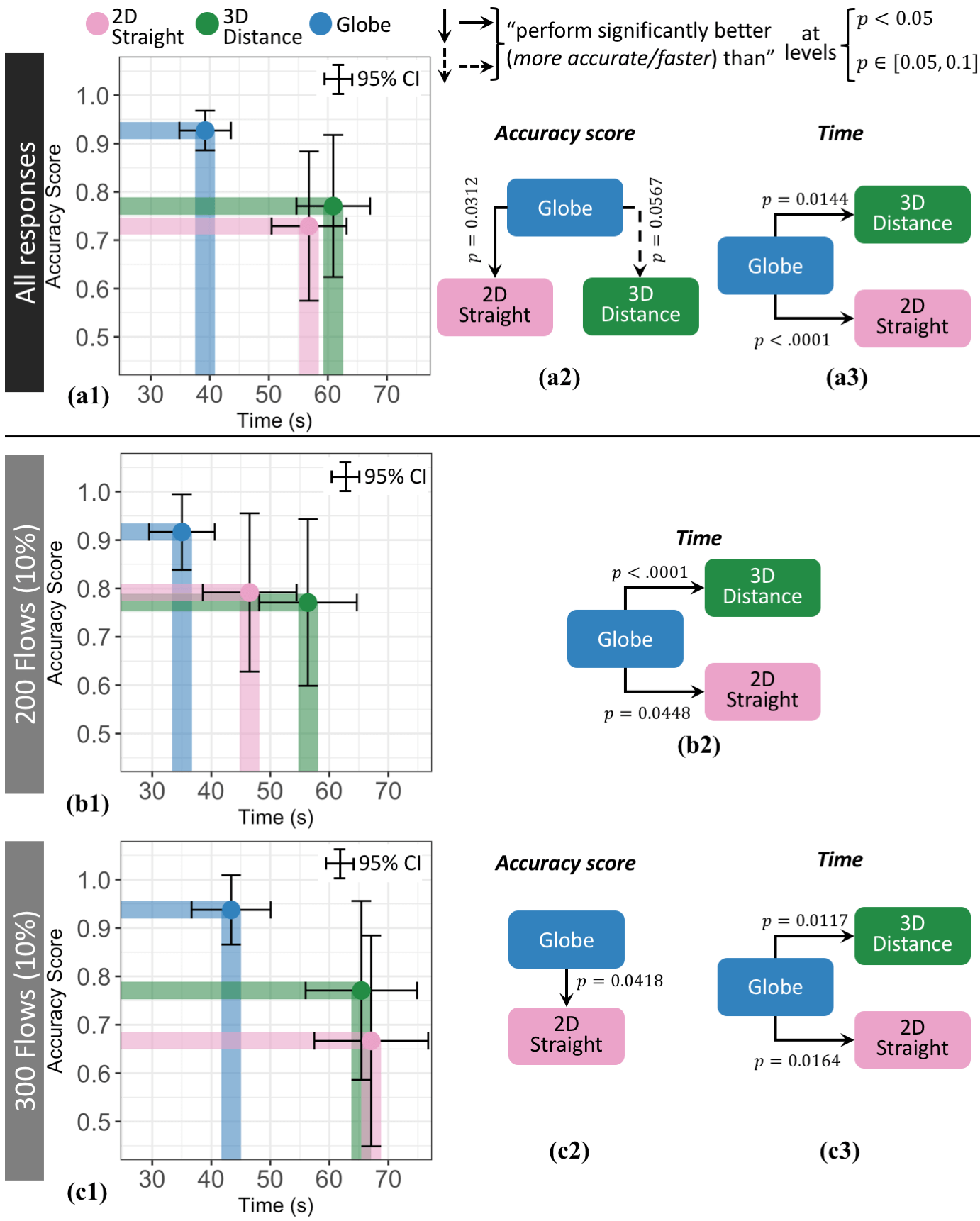}
	\caption{Study 3: (a1, b1, c1) Average performance with 95\% confidence interval, (a2, a3, b2, c2, c3) graphical depiction of results of pairwise post-hoc test.}
	\label{fig:third-study-result-all}
\end{figure}

The same statistical analysis methods were used for accuracy score and $log$ transformed responding time. The Friedman test revealed a statistically significant effect of visualisations on accuracy ($\chi^2(2) = 7.79, p = .0203$). Fig.~\ref{fig:third-study-result-all}~(a1) shows \textit{Globe} (avg. 0.93) was significantly more accurate than \textit{3D distance} (avg. 0.77) and \textit{2D straight} (avg. 0.73). The ANOVA analysis also showed significant effect of visualisations on time ($\chi^2(2) = 11.82, p = .0027$). \textit{Globe} (avg. 39.2s) again was significantly faster than \textit{3D distance} (avg. 60.9s) and \textit{2D straight} (avg. 56.8s). While \textit{3D distance} was slightly more accurate than \textit{2D straight} this was not statistically significant.

By difficulty condition the Friedman test revealed significant effect for accuracy score:\\
\noindent\textbf{\textit{200 flows (10\%)}:} $\chi^2(2) = 2.47, p = .2910$. No statistical significance effect was found in this condition of visualisations.\\
\noindent\textbf{\textit{300 flows (10\%)}:} $\chi^2(2) = 6.26, p = .0437$. \textit{Globe} (avg. 0.94) was significantly more accurate than \textit{2D straight} (avg. 0.67). No significant difference between \textit{3D distance} (avg. 0.77) and other flow maps.

By difficulty condition the ANOVA analysis revealed significant effect for time:\\
\noindent\textbf{\textit{200 flows (10\%)}:} $\chi^2(2) = 13.73, p = .0010$. \textit{Globe} (avg. 35.0s) was significantly faster than both \textit{2D straight} (avg. 46.5s) and \textit{3D distance} (avg. 56.4s).\\ 
\noindent\textbf{\textit{300 flows (10\%)}:} $\chi^2(2) = 8.72, p = .0128$. \textit{Globe} (avg. 43.4s) again was significantly faster than both \textit{2D straight} (avg. 67.1s) and \textit{3D distance} (avg. 65.4s).

\begin{figure}
	\centering
	\includegraphics[width=0.5\textwidth]{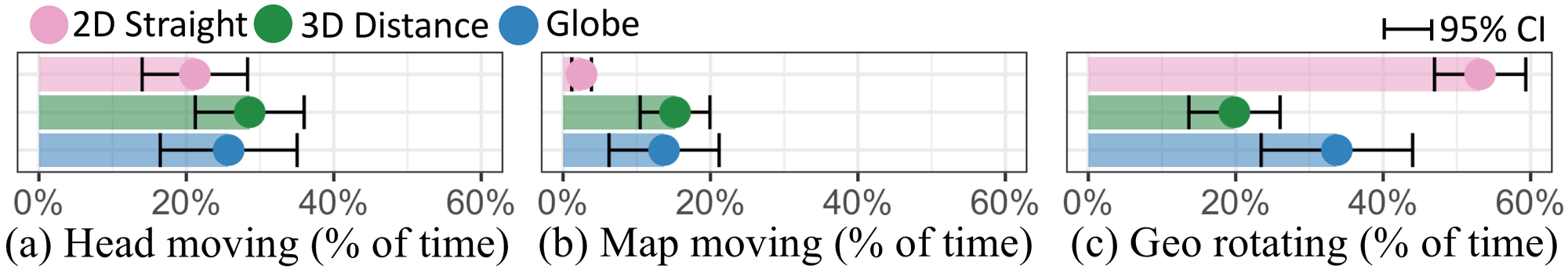}
	\caption{Study 3: Interaction time percentage with 95\% confidence interval}
	\label{fig:third-interaction-percentage}
\end{figure}

\noindent\textbf{Interactions}:
The percentage time difference between interactions per user is demonstrated in Fig.~\ref{fig:third-interaction-percentage}. The Friedman test was used to analyse the relationship between interactions and visualisations.
In \textit{head movement}, there is no significant difference among the visualisations. In \textit{map movement}, \textit{3D distance} $\approx$ \textit{globe} $>$ \textit{2D straight} ($\approx$ means no statistical significance found between two visualisations). In \textit{geo-rotation}, \textit{2D straight} $>$ \textit{3D distance} ($p<.0001$), \textit{2D straight} $>$ \textit{globe} ($p=.0637$) and \textit{globe} $>$ \textit{3D distance} ($p=.0637$).
In \textit{\textit{2D straight}}, \textit{geo-rotation} $>$ head movement $>$ map movement (all $p<.05$). In \textit{\textit{3D distance}}, no statistical significance found among different interactions. In \textit{globe}, \textit{geo-rotation} $>$ map movement ($p=.0216$).

\begin{figure}
	\centering
	\includegraphics[width=0.40\textwidth]{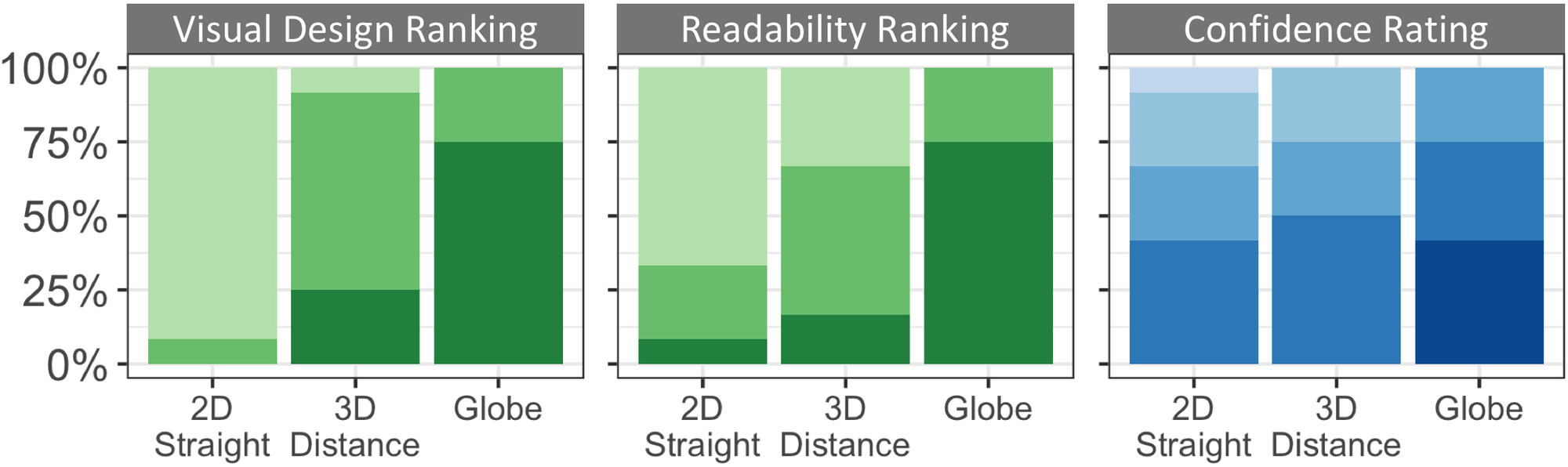}
	\caption{Study 3: Participants preference ranking (\setlength{\fboxsep}{1.5pt}\colorbox{third_first}{\textcolor{white}{$1^{st}$}}, \setlength{\fboxsep}{1pt}\colorbox{third}{\textcolor{black}{$2^{nd}$}} and \setlength{\fboxsep}{1pt}\colorbox{second_fourth}{$3^{rd}$}) and confidence rating (from \setlength{\fboxsep}{1pt}\colorbox{first_conf}{\strut \textcolor{white}{fully confident}} to \setlength{\fboxsep}{2pt}\colorbox{last_conf}{not confident at all}).}
	\label{fig:third-rankings}
\end{figure}

\noindent\textbf{User preference}:
Participant ranking for each of the three visualisations by percentage of respondents is shown by colour (see Fig.~\ref{fig:third-rankings}). For \emph{visual design}, the Friedman test revealed a significant effect of visualisations on preference ($\chi^2(2)=17.17, p = .0001$). Both \textit{globe} (75\% voting it the best) and \textit{3D distance} (25\% voting it the best) were preferred to \textit{2D straight} (0\% voting it as the best) with all $p < .05$. For \emph{readability}, the Friedman test revealed a significant effect of visualisation on preference ($\chi^2(2)=11.17, p = .0038$). \textit{Globe} (75\% voting it the best) was preferred to \textit{2D straight} (8.33\% voting it the best) at significant level $p=.0030$ and \textit{3D distance} (16.67\% voting it the best) at significance level $p=.0638$. As in Study 2, a five-point-Likert scale was used for rating participants' confidence (see Fig.~\ref{fig:third-rankings}). The Friedman test revealed a significant effect of visualisations on confidence ($\chi^2(2)=5.19, p = .07463$). Participants felt more confident with \textit{globe} than \textit{2D straight} at significance level $p=.0954$.

\subsection{Key Findings}
The main finding of Study 3 was confirmation that the findings of Study 2 extend to larger data sets. We found that \textit{globe} was more accurate and faster than \textit{2D straight} and \textit{3D distance} for larger datasets. Overall, \textit{globe} was the preferred visualisation and again participants tended to use \textit{geo-rotation} more than other interactions with both \textit{2D straight} and \textit{globe}. There was some evidence that even with \textit{geo-rotation} \textit{3D distance} scaled better to larger data sets than \textit{2D straight} but this was not statistically significant.

We were surprised by the performance of \textit{globe}, as viewers can only see half of the globe at a time. We therefore investigated performance on items where the OD flows were more than 120\textdegree \ apart. Again we found that in both Studies 2 and 3, performance was better with the \textit{globe} than the other two representations.

\hyphenation{Maps-Link}
\section{Conclusion}
\label{sec:conclusion}

The current paper significantly extends our understanding of how to visualise spatially embedded data in modern immersive environments by systematically investigating and evaluating different 2D and 3D representations for OD flow maps.
We have conducted the first investigation and empirical evaluation of OD flow map visualisation with a modern head-tracked binocular VR HMD. We have found strong evidence that 2D OD flow maps are \textit{not} the best way to show origin-destination flow in such  an environment, and that the use of 3D flow maps can allow viewers to resolve overlapping flows by changing the relative position of the head and object.  However, the particular 3D design choices of the visualisation have a significant effect, for example, encoding flow height to distance was clearly better than to quantity while our most novel use of 3D space, \emph{MapLink}, had the worst performance.

We found that for global flows, the most accurate and preferred representation was a 3D globe with  raised flows whose height is proportional to the flow distance, while for regional flow data the best view would be a flat map with distance-proportional raised 3D flows. We found that accuracy of a standard flat flow map with straight lines could be significantly improved by the use of \emph{geo-rotation}, in which the user can interactively reposition the centre of the map. Nonetheless, the 3D representations were still more accurate and preferred. 

Our findings suggest that globes are preferrable for visualising global OD flow data in \deleted{an interactive VR or MR setting} \added{immersive environments}. For regional OD data, flow should be shown using 3D flows with heights proportional to flow distance. 
Further work could include testing additional encodings and additional tasks including collaborative tasks in multiuser immersive environments.  This study used VR HMDs, as these currently offer the best field-of-view.  The results should be applicable to improved AR headsets as they become available but this should also be tested.

\acknowledgments{
This research was supported under Australian Research Council’s    Discovery Projects funding scheme (project number DP180100755). Data61, CSIRO (formerly NICTA) is funded by the Australian Government through the Department of Communications and the Australian Research Council through the ICT Centre for Excellence Program. We would like to thank all of our user study participants for their time and feedback. We would also like to thank the reviewers for their valuable comments.}

\bibliographystyle{abbrv-doi}

\bibliography{template}

\begin{thebibliography}{10}

\bibitem{Abel:2014iz}
G.~J. Abel and N.~Sander.
\newblock {Quantifying Global International Migration Flows}.
\newblock {\em Science}, 343(6178):1520--1522, 2014.

\bibitem{bach2017immersive}
B.~Bach, M.~Cordeil, T.~Dwyer, B.~Lee, B.~Saket, A.~Endert, C.~Collins, and
  S.~Carpendale.
\newblock Immersive analytics: Exploring future visualization and interaction
  technologies for data analytics.

\bibitem{bach2017towards}
B.~Bach, N.~H. Riche, C.~Hurter, K.~Marriott, and T.~Dwyer.
\newblock Towards unambiguous edge bundling: Investigating confluent drawings
  for network visualization.
\newblock {\em IEEE transactions on visualization and computer graphics},
  23(1):541--550, 2017.

\bibitem{boyandin2011flowstrates}
I.~Boyandin, E.~Bertini, P.~Bak, and D.~Lalanne.
\newblock {Flowstrates: An Approach for Visual Exploration of Temporal
  Origin-Destination Data}.
\newblock {\em Computer Graphics Forum}, 30(3):971--980, 2011.

\bibitem{Buchin:2011fk}
K.~Buchin, B.~Speckmann, and K.~Verbeek.
\newblock {Flow Map Layout via Spiral Trees}.
\newblock {\em IEEE Transactions on Visualization and Computer Graphics},
  17(12):2536--2544, 2011.

\bibitem{buschmann2012challenges}
S.~Buschmann, M.~Trapp, and J.~D{\"o}llner.
\newblock Challenges and approaches for the visualization of movement
  trajectories in 3d geovirtual environments.
\newblock In {\em Proc. GIScience Workshop on GeoVisual Analytics--Time to
  Focus on Time}, vol.~11, 2012.

\bibitem{Butscher:2018ct}
S.~Butscher, S.~Hubenschmid, J.~M\"{u}ller, J.~Fuchs, and H.~Reiterer.
\newblock Clusters, trends, and outliers: How immersive technologies can
  facilitate the collaborative analysis of multidimensional data.
\newblock In {\em Proceedings of the 2018 CHI Conference on Human Factors in
  Computing Systems}, CHI'18, pp. 90:1--12. ACM, 2018.

\bibitem{Immersive15}
T.~Chandler, M.~Cordeil, T.~Czauderna, T.~Dwyer, J.~Glowacki, C.~Goncu,
  M.~Klapperstueck, K.~Klein, K.~Marriott, F.~Schreiber, and E.~Wilson.
\newblock Immersive analytics.
\newblock In {\em Proceedings of 2015 Big Data Visual Analytics (BDVA)}, pp.
  1--8, 2015.

\bibitem{chicagoAreaTransportStudy}
{Chicago Area Transportation Study}.
\newblock {Final Report, Vol I, Survey Findings}, 1959.

\bibitem{Collins:2007ir}
C.~Collins and S.~Carpendale.
\newblock {VisLink: Revealing Relationships Amongst Visualizations}.
\newblock {\em IEEE Transactions on Visualization and Computer Graphics},
  13(6):1192--1199, 2007.

\bibitem{cordeil2017imaxes}
M.~Cordeil, A.~Cunningham, T.~Dwyer, B.~H. Thomas, and K.~Marriott.
\newblock Im{A}xes: Immersive axes as embodied affordances for interactive
  multivariate data visualisation.
\newblock In {\em Proceedings of the 30th Annual ACM Symposium on User
  Interface Software and Technology}, pp. 71--83. ACM, 2017.

\bibitem{Cordeil:2016io}
M.~Cordeil, T.~Dwyer, K.~Klein, B.~Laha, K.~Marriott, and B.~H. Thomas.
\newblock {Immersive Collaborative Analysis of Network Connectivity: CAVE-style
  or Head-Mounted Display?}
\newblock {\em IEEE Transactions on Visualization and Computer Graphics},
  23(1):441--450, 2016.

\bibitem{Cox:1995fp}
K.~C. Cox and S.~G. Eick.
\newblock {Case study: 3D displays of Internet traffic}.
\newblock In {\em Proceedings of Visualization 1995 Conference}, pp. 129--131.
  IEEE, 1995.

\bibitem{Cox:1996gv}
K.~C. Cox, S.~G. Eick, and T.~He.
\newblock {3D geographic network displays}.
\newblock {\em ACM SIGMOD Record}, 25(4):50--54, 1996.

\bibitem{debiasi2014supervised}
A.~Debiasi, B.~Sim{\~o}es, and R.~De~Amicis.
\newblock Supervised force directed algorithm for the generation of flow maps.
\newblock 2014.

\bibitem{dent2008cartography}
B.~D. Dent, J.~S. Torguson, and T.~W. Hodler.
\newblock {\em Cartography: Tematic Map Design}.
\newblock Mc-Graw-Hill Higher Education, 2008.

\bibitem{dubel20142d}
S.~D{\"u}bel, M.~R{\"o}hlig, H.~Schumann, and M.~Trapp.
\newblock 2{D} and 3{D} presentation of spatial data: A systematic review.
\newblock In {\em Proceedings of IEEE VIS International Workshop on 3DVis
  (3DVis)}, pp. 11--18. IEEE, 2014.

\bibitem{dwyer2016ImmersiveAnalytics}
T.~Dwyer, N.~H. Riche, K.~Klein, W.~Stuerzlinger, and B.~Thomas.
\newblock {Immersive Analytics (Dagstuhl Seminar 16231)}.
\newblock {\em Dagstuhl Reports}, 6(6):1--9, 2016.

\bibitem{Eick96aspectsof}
S.~G. Eick.
\newblock Aspects of network visualization.
\newblock {\em IEEE Computer Graphics and Applications}, 16:69--72, 1996.

\bibitem{elsayed2016situated}
N.~A. ElSayed, B.~H. Thomas, K.~Marriott, J.~Piantadosi, and R.~T. Smith.
\newblock Situated analytics: Demonstrating immersive analytical tools with
  augmented reality.
\newblock {\em Journal of Visual Languages \& Computing}, 36:13--23, 2016.

\bibitem{Feiner:1993ip}
S.~Feiner, B.~Macintyre, and D.~Seligmann.
\newblock {Knowledge-based augmented reality}.
\newblock {\em Communications of the ACM}, 36(7):53--62, 1993.

\bibitem{field2012discovering}
A.~Field, J.~Miles, and Z.~Field.
\newblock {\em Discovering statistics using R}.
\newblock Sage publications, 2012.

\bibitem{gilbert2005cattle}
M.~Gilbert, A.~Mitchell, D.~Bourn, J.~Mawdsley, R.~Clifton-Hadley, and W.~Wint.
\newblock Cattle movements and bovine tuberculosis in great britain.
\newblock {\em Nature}, 435(7041):491, 2005.

\bibitem{guo2007visual}
D.~Guo.
\newblock Visual analytics of spatial interaction patterns for pandemic
  decision support.
\newblock {\em International Journal of Geographical Information Science},
  21(8):859--877, 2007.

\bibitem{hagerstrand1970people}
T.~H{\"a}gerstrand.
\newblock What about people in regional science?
\newblock In {\em Papers of the Regional Science Association}, vol.~24, pp.
  6--21. Springer, 1970.

\bibitem{Hollander:1999ns}
M.~Hollander and D.~A. Wolfe.
\newblock {\em {Nonparametric Statistical Methods}}.
\newblock New York: Wiley-Interscience, 1999.

\bibitem{Holten:2011fp}
D.~Holten, P.~Isenberg, J.~J. Van~Wijk, and J.-D. Fekete.
\newblock {An extended evaluation of the readability of tapered, animated, and
  textured directed-edge representations in node-link graphs}.
\newblock In {\em 2011 IEEE Pacific Visualization Symposium (PacificVis)}, pp.
  195--202. IEEE, 2011.

\bibitem{Holten:2009directed}
D.~Holten and J.~J. van Wijk.
\newblock A user study on visualizing directed edges in graphs.
\newblock In {\em Proceedings of the SIGCHI Conference on Human Factors in
  Computing Systems (CHI)}, pp. 2299--2308. ACM, 2009.

\bibitem{Hothorn:2008fy}
T.~Hothorn, K.~Hornik, M.~A. van~de Wiel, and A.~Zeileis.
\newblock {Implementing a Class of Permutation Tests: The coin Package}.
\newblock {\em Journal of Statistical Software}, 28(1):1--23, 2008.

\bibitem{howell2012statistical}
D.~C. Howell.
\newblock {\em Statistical methods for psychology}.
\newblock Cengage Learning, 2012.

\bibitem{huang2007using}
W.~Huang.
\newblock Using eye tracking to investigate graph layout effects.
\newblock In {\em Proceedings of 2007 6th International Asia-Pacific Symposium
  on Visualization (APVIS'07)}, pp. 97--100. IEEE, 2007.

\bibitem{huang2014larger}
W.~Huang, P.~Eades, and S.-H. Hong.
\newblock Larger crossing angles make graphs easier to read.
\newblock {\em Journal of Visual Languages \& Computing}, 25(4):452--465, 2014.

\bibitem{Huang2017AGesture}
Y.-J. Huang, T.~Fujiwara, Y.-X. Lin, W.-C. Lin, and K.-L. Ma.
\newblock A gesture system for graph visualization in virtual reality
  environments.
\newblock In {\em Proceedings of 2017 IEEE Pacific Visualization Symposium
  (PacificVis)}, pp. 41--45, 2017.

\bibitem{itoh2016ImmersiveAnalytics}
T.~Itoh, K.~Marriott, F.~Schreiber, and U.~W\"ossner.
\newblock { Immersive Analytics: A new multidisciplinary initiative to explore
  future interaction technologies for data analytics}.
\newblock {\em NII Shonan Meeting Report}, (2016-2):1--9, 2016.

\bibitem{jenny2017force}
B.~Jenny, D.~M. Stephen, I.~Muehlenhaus, B.~E. Marston, R.~Sharma, E.~Zhang,
  and H.~Jenny.
\newblock Force-directed layout of origin-destination flow maps.
\newblock {\em International Journal of Geographical Information Science},
  31(8):1521--1540, 2017.

\bibitem{Jenny:2017ci}
B.~Jenny, D.~M. Stephen, I.~Muehlenhaus, B.~E. Marston, R.~Sharma, E.~Zhang,
  and H.~Jenny.
\newblock Design principles for origin-destination flow maps.
\newblock {\em Cartography and Geographic Information Science}, 45(1):62--75,
  2018.

\bibitem{kadmon1971komplot}
N.~Kadmon.
\newblock Komplot ``do-it-yourself'' computer cartography.
\newblock {\em The Cartographic Journal}, 8(2):139--144, 1971.

\bibitem{kaya20143d}
E.~Kaya, M.~T. Eren, C.~Doger, and S.~Balcisoy.
\newblock Do 3d visualizations fail? an empirical discussion on 2d and 3d
  representations of the spatio-temporal data.
\newblock In {\em Proceedings of EURASIA GRAPHICS}, 2014.

\bibitem{kern1969Mapit}
R.~Kern and G.~Rushton.
\newblock Mapit: A computer program for production of flow maps, dot maps and
  graduated symbol maps.
\newblock {\em The Cartographic Journal}, 6(2):131--137, 1969.

\bibitem{koylu2017design}
C.~Koylu and D.~Guo.
\newblock Design and evaluation of line symbolizations for origin--destination
  flow maps.
\newblock {\em Information Visualization}, 16(4):309--331, 2017.

\bibitem{Kwon:2016go}
O.-H. Kwon, C.~Muelder, K.~Lee, and K.-L. Ma.
\newblock {A Study of Layout, Rendering, and Interaction Methods for Immersive
  Graph Visualization}.
\newblock {\em IEEE Transactions on Visualization and Computer Graphics},
  22(7):1802--1815, 2016.

\bibitem{munzner2014visualization}
T.~Munzner.
\newblock {\em Visualization analysis and design}.
\newblock CRC {P}ress, 2014.

\bibitem{munzner1996visualizing}
T.~Munzner, E.~Hoffman, K.~Claffy, and B.~Fenner.
\newblock Visualizing the global topology of the {MB}one.
\newblock In {\em Proceedings IEEE Symposium on Information Visualization'96
  (INFOVIS)}, pp. 85--92. IEEE, 1996.

\bibitem{Nocaj:2013gd}
A.~Nocaj and U.~Brandes.
\newblock {Stub Bundling and Confluent Spirals for Geographic Networks}.
\newblock In {\em Graph Drawing}, pp. 388--399. Springer International
  Publishing, 2013.

\bibitem{paci2009knowledge}
R.~Paci and S.~Usai.
\newblock Knowledge flows across {E}uropean regions.
\newblock {\em The Annals of Regional Science}, 43(3):669--690, 2009.

\bibitem{Phan:2005cn}
D.~Phan, L.~Xiao, R.~Yeh, and P.~Hanrahan.
\newblock {Flow map layout}.
\newblock In {\em Proceedings of IEEE Symposium on Information Visualization.
  INFOVIS 2005.}, pp. 219--224, 2005.

\bibitem{purchase1995validating}
H.~C. Purchase, R.~F. Cohen, and M.~James.
\newblock Validating graph drawing aesthetics.
\newblock In {\em International Symposium on Graph Drawing}, pp. 435--446.
  Springer, 1995.

\bibitem{rae2009spatial}
A.~Rae.
\newblock From spatial interaction data to spatial interaction information?
  geovisualisation and spatial structures of migration from the 2001 {UK}
  census.
\newblock {\em Computers, Environment and Urban Systems}, 33(3):161--178, 2009.

\bibitem{riche2012exploring}
N.~H. Riche, T.~Dwyer, B.~Lee, and S.~Carpendale.
\newblock Exploring the design space of interactive link curvature in network
  diagrams.
\newblock In {\em Proceedings of the International Working Conference on
  Advanced Visual Interfaces}, pp. 506--513. ACM, 2012.

\bibitem{Robinson:1955hz}
A.~H. Robinson.
\newblock {The 1837 Maps of Henry Drury Harness}.
\newblock {\em The Geographical Journal}, 121(4):440--450, 1955.

\bibitem{Robinson:1967cj}
A.~H. Robinson.
\newblock {The Thematic Maps of Charles Joseph Minard}.
\newblock {\em Imago Mundi}, 21:95--108, 1967.

\bibitem{scheepens2011composite}
R.~Scheepens, N.~Willems, H.~Van~de Wetering, G.~Andrienko, N.~Andrienko, and
  J.~J. Van~Wijk.
\newblock Composite density maps for multivariate trajectories.
\newblock {\em IEEE Transactions on Visualization and Computer Graphics},
  17(12):2518--2527, 2011.

\bibitem{stephen2017automated}
D.~M. Stephen and B.~Jenny.
\newblock Automated layout of origin--destination flow maps: {US}
  county-to-county migration 2009--2013.
\newblock {\em Journal of Maps}, 13(1):46--55, 2017.

\bibitem{Sun:2018db}
S.~Sun.
\newblock {A spatial one-to-many flow layout algorithm using triangulation,
  approximate Steiner trees, and path smoothing}.
\newblock {\em Cartography and Geographic Information Science}, 11(1):1--17,
  2018.

\bibitem{Tobler:1981kw}
W.~Tobler.
\newblock {Depicting Federal Fiscal Transfers}.
\newblock {\em The Professional Geographer}, 33(4):419--422, 1981.

\bibitem{Tobler:1987kn}
W.~R. Tobler.
\newblock {Experiments In Migration Mapping By Computer}.
\newblock {\em Cartography and Geographic Information Science}, 14(2):155--163,
  1987.

\bibitem{van2014multivariate}
S.~van~den Elzen and J.~J. van Wijk.
\newblock Multivariate network exploration and presentation: From detail to
  overview via selections and aggregations.
\newblock {\em IEEE Transactions on Visualization and Computer Graphics},
  20(12):2310--2319, 2014.

\bibitem{Vrotsou:2017im}
K.~Vrotsou, G.~Fuchs, N.~Andrienko, and G.~Andrienko.
\newblock {An Interactive Approach for Exploration of Flows Through
  Direction-Based Filtering}.
\newblock {\em Journal of Geovisualization and Spatial Analysis}, 1(1-2):205,
  2017.

\bibitem{ware2005reevaluating}
C.~Ware and P.~Mitchell.
\newblock Reevaluating stereo and motion cues for visualizing graphs in three
  dimensions.
\newblock In {\em Proceedings of the 2nd Symposium on Applied Perception in
  Graphics and Visualization}, pp. 51--58. ACM, 2005.

\bibitem{Willett:2015fv}
W.~Willett, B.~Jenny, T.~Isenberg, and P.~Dragicevic.
\newblock {Lightweight Relief Shearing for Enhanced Terrain Perception on
  Interactive Maps}.
\newblock In {\em Proceedings of the 33rd Annual ACM Conference on Human
  Factors in Computing Systems (CHI)}, pp. 3563--3572. ACM, 2015.

\bibitem{wittick1976}
R.~I. Wittick.
\newblock A computer system for mapping and analyzing transportation networks.
\newblock {\em Southeastern Geographer}, 16(1):74--81, 1976.

\bibitem{wong2003edgelens}
N.~Wong, S.~Carpendale, and S.~Greenberg.
\newblock Edgelens: An interactive method for managing edge congestion in
  graphs.
\newblock In {\em IEEE Symposium on Information Visualization, 2003. INFOVIS
  2003}, pp. 51--58. IEEE, 2003.

\bibitem{wood2010visualisation}
J.~Wood, J.~Dykes, and A.~Slingsby.
\newblock Visualisation of origins, destinations and flows with od maps.
\newblock {\em The Cartographic Journal}, 47(2):117--129, 2010.

\bibitem{Yang:2017cy}
Y.~Yang, T.~Dwyer, S.~Goodwin, and K.~Marriott.
\newblock {Many-to-Many Geographically-Embedded Flow Visualisation: An
  Evaluation}.
\newblock {\em IEEE Transactions on Visualization and Computer Graphics},
  23(1):411--420, 2017.

\bibitem{Yang:2018mg}
Y.~Yang, B.~Jenny, T.~Dwyer, K.~Marriott, H.~Chen, and M.~Cordeil.
\newblock {Maps and Globes in Virtual Reality}.
\newblock {\em Computer Graphics Forum}, 37(3):427--438, 2018.

\bibitem{Zhang:2016}
M.-J. Zhang, J.~Li, and K.~Zhang.
\newblock An immersive approach to the visual exploration of geospatial network
  datasets.
\newblock In {\em Proceedings of the 15th ACM SIGGRAPH Conference on
  Virtual-Reality Continuum and Its Applications in Industry}, VRCAI '16, pp.
  381--390. ACM, 2016.

\bibitem{Zhang:2018jo}
M.-J. Zhang, K.~Zhang, J.~Li, and Y.-N. Li.
\newblock {Visual Exploration of 3D Geospatial Networks in a Virtual Reality
  Environment}.
\newblock {\em The Computer Journal}, 61(3):447--458, 2018.

\end{thebibliography}
\end{document}